\documentclass{aa}  

\usepackage{graphicx}
\usepackage{txfonts}
\def\PLUTO{{\sc pluto}}

\newcommand\rs[1]{_\mathrm{#1}}

\newcommand{\casa}{Cas~A}

\begin{document} 

    \title{Filamentary ejecta network in Cassiopeia~A reveals fingerprints of the supernova explosion mechanism}

   \author{S.\,Orlando\inst{1}
      \and H.-T.\,Janka\inst{2}
      \and A.\,Wongwathanarat\inst{2}
      \and D.\,Dickinson\inst{3}
      \and D.\,Milisavljevic\inst{3}
      \and M.\,Miceli\inst{4,1}
      \and F.\,Bocchino\inst{1}
      \and T.\,Temim\inst{5}
      \and I.\,De Looze\inst{6}
      \and D.\,Patnaude\inst{7}
  }

\institute{INAF -- Osservatorio Astronomico di Palermo, Piazza del Parlamento 1, I-90134 Palermo, Italy\\
\email{salvatore.orlando@inaf.it}
\and Max-Planck-Institut f\"ur Astrophysik, Karl-Schwarzschild-Str. 1, D-85748 Garching, Germany
\and Department of Physics and Astronomy, Purdue University, 525 Northwestern Avenue, West Lafayette, IN 47907, USA
\and Dip. di Fisica e Chimica, E. Segr\'e, Universit\`a degli Studi di Palermo, Piazza del Parlamento 1, 90134 Palermo, Italy
\and Princeton University, 4 Ivy Ln, Princeton, NJ 08544, USA
\and Sterrenkundig Observatorium, Ghent University, Krijgslaan 281 - S9, B-9000 Gent, Belgium
\and Smithsonian Astrophysical Observatory, 60 Garden Street, Cambridge, MA 02138, USA
             }

   \date{Received XXX; accepted XXX}

 
  \abstract
   {Recent observations with the James Webb Space Telescope (JWST) have revealed unprecedented details of an intricate filamentary structure of unshocked ejecta within the young supernova remnant (SNR) Cassiopeia A (\casa), offering new insights into the mechanisms governing supernova (SN) explosions and the subsequent evolution of ejecta.}
   {We aim to investigate the origin and evolution of the newly discovered web-like network of ejecta filaments in \casa. Our specific objectives are: (i) to characterize the three-dimensional (3D) structure and kinematics of the filamentary network and (ii) to identify the physical mechanisms responsible for its formation.}
   {We performed high-resolution, 3D hydrodynamic (HD) and magneto-hydrodynamic (MHD) simulations to model the evolution of a neutrino-driven SN from the explosion to its remnant with the age of 1000~years. The initial conditions, set shortly after the shock breakout at the stellar surface, are based on a 3D neutrino-driven SN model that closely matches the basic properties of \casa. 
   }
   {We found that the magnetic field has little impact on the evolution of unshocked ejecta, so we focused most of the analysis on the HD simulations. A web-like network of ejecta filaments, with structures compatible with those observed by JWST (down to scales $\approx 0.01$~pc), naturally forms during the SN explosion. The filaments result from the combined effects of processes occurring soon after the core collapse, including the expansion of neutrino-heated bubbles formed within the first second after the explosion, hydrodynamic instabilities triggered during the blast propagation through the stellar interior, and the Ni-bubble effect following the shock breakout. The interaction of the reverse shock with the ejecta progressively disrupts the filaments through the growth of hydrodynamic instabilities. By around 700~years, the filamentary network becomes unobservable.} 
   {According to our models, the filaments observed by JWST in \casa\ most likely preserve a "memory" of the early explosion conditions, reflecting the processes active during and immediately after the SN event. Notably, a filamentary network closely resembling that observed in \casa\ is naturally produced by a neutrino-driven SN explosion.}

               
\titlerunning{Filamentary Ejecta Network in Cassiopeia~A Reveals Fingerprints of the Explosion Mechanism}

\authorrunning{S. Orlando et~al.}

   \maketitle
%

\section{Introduction}

Supernova remnants (SNRs) are unique laboratories for probing the physics of stellar explosions. Among these, Cassiopeia~A (\casa), at an estimated age of about 350 years, represents the youngest known core-collapse SNR in our Galaxy (e.g., \citealt{2001AJ....122..297T, 2006ApJ...645..283F}). Its proximity to Earth ($\sim 3.4$~kpc; \citealt{1995ApJ...440..706R}) and its youth make \casa\ an exceptional target for studying the early evolution of supernova (SN) ejecta, offering invaluable insights into the physical processes shaping its structure and morphology.
 
The internal structure of SN ejecta holds crucial information about both the explosion mechanism and the properties of the progenitor star. Studies of \casa\ have shown that its ejecta possess complex morphological features, including high-velocity knots, jet-like structures, and large-scale asymmetries (e.g., \citealt{1996ApJ...470..967F, 2003ApJ...597..347L, 2003ApJ...597..362H, 2010ApJ...725.2038D, 2013ApJ...772..134M, 2014Natur.506..339G, 2015Sci...347..526M}). Of particular interest are the spatial distributions of elements synthesized during the SN, such as $^{56}$Ni and $^{44}$Ti and their decay products, as they may still reflect the key physical processes that governed the explosion even hundreds of years after core collapse. These dominant asymmetries in the ejecta structures, along with evidence that \casa\ is still expanding within the progenitor star's wind-blown cavity (e.g., \citealt{2014ApJ...789....7L}), suggest an asymmetric explosion likely driven by neutrino-heated convection and the standing accretion shock instability (SASI; \citealt{2003ApJ...584..971B, 2012ARNPS..62..407J, 2017hsn..book.1095J, 2018SSRv..214...33B, 2021Natur.589...29B}).

This interpretation is strongly supported by three-dimensional (3D) hydrodynamic (HD) and magneto-hydrodynamic (MHD) models that track the SNR evolution from explosion to its present state (e.g., \citealt{2016ApJ...822...22O, 2017ApJ...842...13W, 2021A&A...645A..66O}). These models show that many observed features of \casa\ arise naturally from stochastic processes such as convective overturn and SASI, which develop within seconds after the core-collapse. As a result, there is growing interest in identifying specific explosion signatures within the ejecta, as these could reveal critical aspects of the physical processes governing the complex phases of core-collapse SNe and probe the conditions during the earliest moments of stellar death.

Recent high-resolution observations of \casa\ with the James Webb Space Telescope (JWST) have opened new avenues for exploring such signatures within the ejecta. These observations offer an unprecedented view of the remnant's interior structure, unveiling a previously unrecognized network of intricately structured filaments interior to \casa’s main shell of reverse-shocked material (\citealt{2024ApJ...965L..27M}; see left panel of Fig.~\ref{whole_remnant}). These features are resolved at scales as fine as $\sim 0.01$~pc, a remarkable achievement given the remnant's extensive $\approx 5$~pc diameter. The filaments exhibit distinct chemical signatures, notably oxygen-rich (O) compositions, and are organized hierarchically, suggesting a complex interplay of physical processes during their formation and evolution. In a companion paper, the JWST data are analyzed to investigate the detailed properties of these filaments, including their chemical composition, kinematics, and spatial distribution, with the goal of linking these observed features to the underlying physical processes that shaped them (Dickinson et al., in preparation).

The origin of the filamentary structures within the innermost ejecta of \casa\ is still uncertain. Various mechanisms have been proposed to explain filaments observed in the mixing regions between forward and reverse shocks in many SNRs, including Rayleigh-Taylor instabilities at the interfaces between different ejecta layers (\citealt{1973MNRAS.161...47G, 1992ApJ...392..118C}), and magnetic field amplification through turbulent dynamo action (\citealt{1996ApJ...465..800J}). However, the filaments recently uncovered by JWST in \casa\ are located interior to the reverse shock, suggesting that these structures may originate from effects different from those previously considered and may possibly be directly linked to the explosion mechanism instead. The specific processes behind these patterns and their roles in forming such intricate filamentary structures remain open questions.

In this paper, we investigate this issue using the 3D HD/MHD model from \cite{2021A&A...645A..66O}, complemented by high-resolution simulations designed to capture ejecta features at scales approaching those observed by JWST. This model traces the evolution of a neutrino-driven SN explosion, replicating key characteristics of \casa\ from the immediate post-core-collapse phase to its current state. In such a way, this approach provides a robust framework to link the observed filaments with specific physical processes active throughout the SN and the remnant evolution. Our study focuses on two main objectives: (i) characterizing the morphological and kinematic properties of the network of filament by comparing the simulated structures with the results from observations, (ii) tracing the ejecta evolution from the initial seconds after the explosion to examine how filamentary features observed today are connected to physical processes and phenomena triggered shortly after core collapse.

Our study, therefore, offers the possibility to infer potential constraints that the filaments structure may impose on the SN explosion parameters, to identify effects that may be missing in our employed explosion model (e.g., connected to the \casa\ ``jets''), and to investigate whether serious conflicts with observational properties of the \casa\ interior occur. Such discrepancies could challenge our current understanding of the core-collapse SN mechanism and highlight areas requiring further refinement in theoretical models.

The paper is organized as follows: Sect.~\ref{sec:model} summarizes our HD/MHD simulation setup and methodology; Sect.~\ref{sec:results} presents the results, focusing on the 3D structure of the unshocked ejecta and their origins tracing back to the core-collapse; finally, Sect.~\ref{sec:conc} summarizes our main conclusions and discusses their broader implications for understanding SN explosion mechanisms and remnant evolution. In Appendix~\ref{app:resolution}, we investigate the effects of the spatial resolution on the results; in Appendix~\ref{app:MHD}, we evaluate the effects of the magnetic field on the structure of unshocked ejecta; in Appendix~\ref{app:multi-media}, we describe the content of online multi-media material, offering interactive and dynamic representations of the
phenomena discussed in the paper.

\section{The 3D HD/MHD model}
\label{sec:model}

The numerical setup used in this study follows the approach detailed in previous works (e.g., \citealt{2021A&A...645A..66O}) and models the evolution of a neutrino-driven SN from the core collapse to its remnant over approximately 1000 years. More specifically, we continue a SN explosion model (W15-2-cw-IIb; \citealt{2017ApJ...842...13W}), selected for its close match to the basic properties of \casa, by subsequent HD or MHD simulations to investigate the remnant expansion and its interaction with the circumstellar medium (CSM; \citealt{2022A&A...666A...2O}). A detailed description of the numerical setup can be found in \cite{2017ApJ...842...13W} for the SN and \cite{2021A&A...645A..66O, 2022A&A...666A...2O} for the SNR. Below, we provide a summary of its key features and the physical processes included in the model.

The SN model begins with a zero-age main sequence progenitor of $15\,M_{\odot}$ (model W15; \citealt{1995ApJS..101..181W}), which has undergone extensive mass loss, leaving only approximately $0.3\,M_{\odot}$ of its hydrogen (H) envelope (\citealt{2017ApJ...842...13W}). This configuration aligns with observational evidence suggesting that \casa\ resulted from a Type IIb SN (\citealt{2008Sci...320.1195K, 2011ApJ...732....3R}). The explosion is driven by neutrino energy deposition, modeled parametrically to produce an explosion energy of $\approx 1.5 \times 10^{51}$~erg and to eject $\approx 3.3\,M_{\odot}$ of stellar debris into the surrounding CSM (see Table~\ref{Tab:model}). While the ejecta mass agrees with observational estimates, the explosion energy is approximately 25\% lower than the inferred value of $\approx 2 \times 10^{51}$~erg  (e.g., \citealt{2003ApJ...597..347L, 2003ApJ...597..362H, 2020ApJ...893...49S}), as the SN model was not specifically fine-tuned to match all basic properties of \casa.

The 3D SN simulation tracks the evolution from approximately 15 milliseconds after the core bounce to about one day post-explosion, capturing key stages of the event, including the shock breakout at the stellar surface, which occurs at around 1500 seconds (\citealt{2017ApJ...842...13W}). The early explosion phase, extending to several seconds after core bounce, incorporates a parametric treatment of neutrino effects as described in \cite{2010ApJ...725L.106W, 2013A&A...552A.126W, 2015A&A...577A..48W}. The subsequent long-term evolution of the explosion includes essential physical processes that shape the dynamics, such as self-gravity, the gravitational influence of the newly formed neutron star, fallback of material, and the Helmholtz equation of state, which accurately captures thermodynamic properties in the dense core regions.

To track explosive nucleosynthesis, the model employs an $\alpha$-network of 11 nuclear species, providing approximate insights into the chemical composition of the ejected material. Asymmetries in the ejecta distribution naturally arise due to stochastic processes, including nonradial HD instabilities. These instabilities occur both during the onset of the explosion (convective overturn and SASI) and at the composition interfaces of the progenitor following the passage of the outward-propagating shock wave (Rayleigh-Taylor, Kelvin-Helmholtz, and Richtmyer-Meshkov instabilities). For further details on the SN model setup and its early evolution, refer to \cite{2010ApJ...725L.106W, 2013A&A...552A.126W, 2015A&A...577A..48W, 2017ApJ...842...13W}.

The output from the 3D SN simulation at $\approx 17.85$~hours after core-collapse served as the initial condition for 3D simulations that track the transition from the SN phase to the SNR phase and the subsequent expansion of the remnant through the CSM (\citealt{2021A&A...645A..66O, 2022A&A...666A...2O}). The CSM environment is modeled as a stellar wind with a density profile proportional to $r^{-2}$, normalized to $0.8$~cm$^{-3}$ at a radius of 2.5~pc (in agreement with \citealt{2014ApJ...789....7L}), and includes an asymmetric, dense circumstellar shell with its densest region positioned on the near side to the northwest (\citealt{2022A&A...666A...2O}). This shell is included to account for significant asymmetries of the reverse shock, which cannot be explained by models assuming a spherically symmetric stellar wind from the progenitor (see \citealt{2022A&A...666A...2O}). Recent JWST observations support the existence of this shell, having revealed evidence of a shocked, dense CSM structure around \casa\ (the so-called ``Green Monster''), consistent with past interactions between the remnant and a dense region of CSM material (\citealt{2024ApJ...965L..27M, 2024ApJ...976L...4D, Orlando2025}). In the MHD simulations, the ambient magnetic field is represented by a ``Parker spiral'' configuration, resulting from the progenitor star's rotation and the corresponding outward flow of the stellar wind (\citealt{1958ApJ...128..664P}; see \citealt{2019A&A...622A..73O} for further details).

The SNR simulations include: (i) energy deposition from radioactive decay, specifically the $^{56}$Ni $\rightarrow$ $^{56}$Co $\rightarrow$ $^{56}$Fe chain, modeled by an internal energy term that assumes local deposition, excluding neutrinos that escape freely, and assuming no $\gamma$-ray leakage from the inner part of the remnant (\citealt{2021A&A...645A..66O}); (ii) deviations from ionization equilibrium, estimated through the maximum ionization age in each cell of the spatial domain (\citealt{2015ApJ...810..168O}); and (iii) deviations from electron-proton temperature equilibration, with initial electron heating at shock fronts up to $kT = 0.3$~keV  (\citealt{2007ApJ...654L..69G}), and post-shock ion/electron temperatures determined via Coulomb collisions (\citealt{2015ApJ...810..168O}). In this paper, we omitted the impact of cosmic ray back-reaction at shock fronts explored in previous studies (\citealt{2016ApJ...822...22O, 2022A&A...666A...2O}). This process primarily influences the dynamics of the forward shock and the structure of the mixing region between the forward and reverse shocks (e.g., \citealt{2012ApJ...749..156O, 2018ApJ...852...84P}). However, it does not significantly affect the evolution of the unshocked ejecta, which is the primary focus of this work.

The SNR simulations were conducted using the \PLUTO\ code (\citealt{2007ApJS..170..228M, 2012ApJS..198....7M}), configured with specialized Riemann solvers for both HD and MHD calculations. Specifically, the linearized Roe Riemann solver was applied for HD simulations, while the HLLD approximate Riemann solver was used for MHD simulations. The 3D setup uses a Cartesian coordinate system $(x,y,z)$.

In this study, we increased the spatial resolution by a factor of 2 compared to previous simulations (\citealt{2021A&A...645A..66O, 2022A&A...666A...2O}), enhancing our ability to capture small-scale features in the ejecta distribution that can be directly compared with the high angular resolution of JWST (\citealt{2024ApJ...965L..27M}). To achieve this, the computational domain was covered by a uniform grid of $(2048)^3$ cells, allowing it to expand as the forward shock propagates outward from shock breakout to 1000 years post-explosion, through a sequence of consecutive remappings (see \citealt{2021A&A...645A..66O} for more detail). As a result, the spatial resolution ranged from approximately $1.2\times 10^{11}$~cm at the start (in a domain spanning from $-1.2 \times 10^{14}$ to $1.2 \times 10^{14}$~cm in all directions) to about $0.005$~pc at 1000~years (within a domain extending from $-5.4$ to $5.4$~pc). At the age of \casa\ ($\approx 350$~years), the spatial resolution achieved is $\sim 0.002$~pc, which is slightly higher than the spatial resolution that JWST can achieve for \casa. 

Table~\ref{Tab:model} summarizes the main parameters of the new high-resolution simulations conducted in this study, which were calculated on a 3D Cartesian grid of $(2048)^3$ cells. These simulations replicate the setup of W15-IIb-sh-HD-1eta-az from \cite{2022A&A...666A...2O}, excluding the feedback of particle acceleration. The study explores variations in spatial resolution by considering different computational grids, and examines the effects of radioactive decay and magnetic fields by selectively disabling each of these processes. For comparison, it also includes the main parameters of the SN simulation described in \cite{2017ApJ...842...13W}. The naming convention for the models is as follows: {\em W15} denotes the progenitor star model adopted (\citealt{1995ApJS..101..181W}); {\em IIb} indicates that the progenitor was stripped of its envelope, resulting in a SN Type IIb (\citealt{2017ApJ...842...13W}); {\em sh} indicates interaction with a circumstellar shell (\citealt{2022A&A...666A...2O}); {\em HD/MHD} specifies whether the simulation is HD or MHD; {\em dec} indicates inclusion of the Ni-bubble effect; and {\em hr} identifies high-resolution simulations ($2048^3$ grid points). In Appendix~\ref{app:resolution}, we present a comparison between two simulations, W15-IIb-sh-HD+dec-hr and W15-IIb-sh-HD+dec (see Table~\ref{Tab:model}), which differ only in their spatial resolution. This comparison aims to assess the model's capability to accurately capture the small-scale structures of the unshocked ejecta.

\begin{table}
\caption{Setup for the simulated models.}
\label{Tab:model}
\begin{center}
\begin{tabular}{llll}
\hline
\hline
SN model & \multicolumn{2}{l}{Parameter}  &  Value  \\
\hline
W15-2-cw-IIb$^{a}$  & \multicolumn{2}{l}{$E\rs{exp}$} &  $1.5$~B$^{b}$  \\
    & \multicolumn{2}{l}{$M\rs{ej}$}  &  $3.3\,M_{\odot}$  \\
    & \multicolumn{2}{l}{$E\rs{exp}/M\rs{ej}$} &  $0.45$~B$/M_{\odot}$  \\ \\ \hline\hline
SNR Model                 & simul. &  rad.  &  Grid\\ 
                          &        &  decay  &  \\ \hline
W15-IIb-sh-HD             & HD     & no     & $(1024)^3$ \\
W15-IIb-sh-HD+dec$^{c}$   & HD     & yes    & $(1024)^3$ \\
W15-IIb-sh-HD+dec-hr      & HD     & yes    & $(2048)^3$ \\
W15-IIb-sh-MHD+dec-hr     & MHD    & yes    & $(2048)^3$ \\
\hline
\end{tabular}
\end{center}
$(a)$ Model presented in \cite{2017ApJ...842...13W};
$(b)$ Where 1 B $ = 10^{51}$~erg;
$(c)$ Model replicating the setup of W15-IIb-sh-HD-1eta-az from \cite{2022A&A...666A...2O}, but without including particle acceleration feedback.
\end{table}

\section{Results}
\label{sec:results}

\begin{table*}
\caption{Masses of ejecta elements (total, shocked, and unshocked) in the W15-IIb-sh-HD+dec-hr model at 351~years.}
\label{Tab:elem}
\begin{center}
\begin{tabular}{llllllll|lll}
\hline
\hline
    & & \multicolumn{5}{c}{W15-IIb-sh-HD+dec-hr} & &  \multicolumn{1}{c}{(a)} & \multicolumn{2}{c}{(b)}  \\ \hline
    & \multicolumn{2}{l}{Total Mass} & \multicolumn{2}{l}{Shocked Mass} & \multicolumn{2}{l}{Unshocked Mass} & $M_{\rm sh}$/ &  Shocked Mass & \multicolumn{2}{l}{Unshocked Mass} \\
    & $M_{\rm tot}/M_{\odot}$ & \% & $M_{\rm sh}/M_{\odot}$ & \% & $M_{\rm ush}/M_{\odot}$ & \% &  $M_{\rm ush}$ & $M_{\rm sh}/M_{\odot}$ & $M_{\rm ush}/M_{\odot}$ & \%  \\
\hline 
ejecta  & 3.3  & -    & 2.5   & -    & 0.79   &  -   & 3.2   & - & $0.47^{+0.47}_{-0.24}$ & - \\
He  &  1.98    & 58   &  1.87 &  72  &  0.11  &  14  &  17   & - & - & - \\
C   &  0.17    & 5.0  & 0.10  &  3.9 &  0.068 &  8.6 &  1.5  & - & - & - \\
O   &  0.59    & 17   & 0.25  &  9.6 &  0.34  &  43  &  0.73 & $[1.80, 2.55]$ &  $0.14 \pm 0.02$ & 30   \\
Ne  &  0.14    & 4.1  & 0.068 &  2.6 &  0.072 &  9.1 &  0.94 & $[0.027, 0.038]$ &  $0.0023\pm 0.0016$ & 0.26 \\
Mg  &  0.038   & 1.1  & 0.015 &  0.57&  0.023 &  2.9 &  0.65 & $[0.007, 0.01]$ &  - & - \\
Si  &  0.027   & 0.79 & 0.010 &  0.38&  0.017 &  2.2 &  0.59 & $[0.038, 0.054]$ &  $0.31 \pm 0.05$ & 60   \\
Ca  &  0.054   & 1.6  & 0.012 &  0.46&  0.042 &  5.4 &  0.29 & - & - & - \\
Ti$^{c}$&0.0028& 0.082& 0.0070& 0.027&  0.0021&  0.26&  0.33 & - & - & - \\
Fe  &  0.093   & 2.7  & 0.029 &  1.1 &  0.064 &  8.1 &  0.45 & $[0.098, 0.14]^d$ & $< 0.07$ & $<15$  \\
X$^{e}$&  0.041& 1.2  & 0.014 &  0.54&  0.027 &  3.4 &  0.52 & - & - & - \\
\hline
\end{tabular}
\end{center}
Notes: The table also includes the ratio of shocked to unshocked masses. The last three columns provide, for comparison, the shocked mass values inferred from Chandra data analysis of \casa\ (\citealt{2012ApJ...746..130H}) and the unshocked mass values inferred from Spitzer data analysis (\citealt{2020ApJ...904..115L}).
$(a)$ Data from \cite{2012ApJ...746..130H}; each element is presented with two values, corresponding to assumed filling factors based on density spike thickness of $2.5^{\prime\prime}$ and $5^{\prime\prime}$;
$(b)$ Data from \cite{2020ApJ...904..115L};
$(c)$ The ejecta mass of $^{44}$Ti is overestimated by a factor $10-20$ due to a reduced network in the original SN model, which included only 9 of 13 $\alpha$ nuclei up to $^{56}$Ni (see \citealt{2017ApJ...842...13W});
$(d)$ The values reported consider the total Fe masses from incomplete Si burning (Fe$_{\rm Si}$) plus from complete Si burning and/or $\alpha$-rich freezout (Fe$_{\rm \alpha}$) (\citealt{2012ApJ...746..130H});
$(e)$ The tracer nucleus $^{56}$X represents Fe-group
species synthesized in neutron-rich environments as, for instance, in neutrino-heated ejecta (see \citealt{2017ApJ...842...13W}). The table lists the mass of $^{56}$Fe, calculated as $M_{\rm Fe} + 0.5 M_{\rm X}$, and the mass of $^{56}$X, estimated as $0.5 M_{\rm X}$.
\end{table*}

The evolution from the SN to the fully developed SNR as described by our models has been detailed in three previous studies. \cite{2017ApJ...842...13W} presented the 3D simulation of the first day of evolution in the adopted neutrino-driven core-collapse SN, showing that the model reproduces key asymmetries observed in \casa\ about a day after shock breakout at the stellar surface. These include the distributions of $^{56}$Ni and $^{44}$Ti, as well as the neutron star kick direction, closely resembling those observed in \casa. This study showed that such features can naturally arise without the need for rapid rotation or jet-driven mechanisms. In \cite{2021A&A...645A..66O}, we extended the SN evolution to its remnant with an age of 2000~years, demonstrating that the large-scale asymmetries from the explosion, interacting with the reverse shock, result in a remnant morphology at the age of \casa\ characterized by ring-like structures, layer inversions, voids, and asymmetric distributions of $^{44}$Ti and $^{56}$Fe, closely matching \casa\ observations and leaving detectable fingerprints up to $\approx 2000$~years post-explosion. In a subsequent study (\citealt{2022A&A...666A...2O}), we showed that some of the enigmatic properties of the reverse shock in \casa, such as its inward or stationary motion in the northwest quadrant (\citealt{2022ApJ...929...57V, 2025arXiv250107708F}), can be explained by the interaction of the remnant with an asymmetric, thin, dense shell (radius $\approx 1.5$~pc, thickness $\approx 0.02$~pc) in the CSM. This shell likely originated from a massive eruption of the progenitor star that occurred $10^4$–$10^5$~years before core collapse. 

In this paper, we further explore the above models to investigate the origin of the intricate network of ejecta filaments revealed by JWST. Since this network lies interior to \casa’s main shell of reverse-shocked material, our analysis focused on the unshocked ejecta, where the original imprint of the SN explosion is possibly best preserved; the evolution of shocked ejecta can be found in \cite{2021A&A...645A..66O}. In Sect.~\ref{sec:res1}, we perform a detailed analysis of the filament network as modeled at \casa’s current age, examining its spatial distribution, velocity structure, and chemical composition, and comparing these model results to JWST observations (\citealt{2024ApJ...965L..27M}; Dickinson et al., in preparation). In Sect.\ref{sec:res2}, we trace the evolution of these structures from the immediate aftermath of core-collapse ($\sim 1$~s after explosion) to \casa’s current age ($\sim 350$~years), with the aim of identifying key physical processes responsible for the formation of this filamentary network. Finally, in Sect.~\ref{sec:res3}, we extend our analysis to an age of 1000~years, to explore the future evolution of the network and to determine approximately how long these features can remain observable.

\subsection{The structure of unshocked ejecta at the age of \casa}
\label{sec:res1}

To facilitate comparison with observations of \casa, we rotated the simulated model around the three axes to align the Ni-rich fingers produced by the original SN model (\citealt{2017ApJ...842...13W}) with the extended Fe-rich regions observed in \casa. The rotation angles applied were $ix = -30^\circ$, $iy = 70^\circ$, and $iz = 10^\circ$ (see \citealt{2021A&A...645A..66O}), a configuration we adopted consistently throughout this study. In this orientation, Earth’s vantage point corresponds to the negative $y$-axis. 

From our analysis of the distribution of unshocked species, we found no significant differences between models with or without magnetic fields (see Table~\ref{Tab:model}). Therefore, in the following, we focused on the results of the run W15-IIb-sh-HD+dec-hr, noting any differences observed in the magnetized case (W15-IIb-sh-MHD+dec-hr) where relevant. In Appendix \ref{app:MHD}, we report a selection of results from run W15-IIb-sh-MHD+dec-hr. The effects of magnetic fields, particularly in shaping the structure of the Green Monster, are discussed in a separate study (\citealt{Orlando2025}). Unless otherwise stated, the time in this study is measured from the core bounce (see \citealt{2017ApJ...842...13W}), also referred to as the time after core-collapse.


   \begin{figure*}
   \centering
   \includegraphics[width=\textwidth]{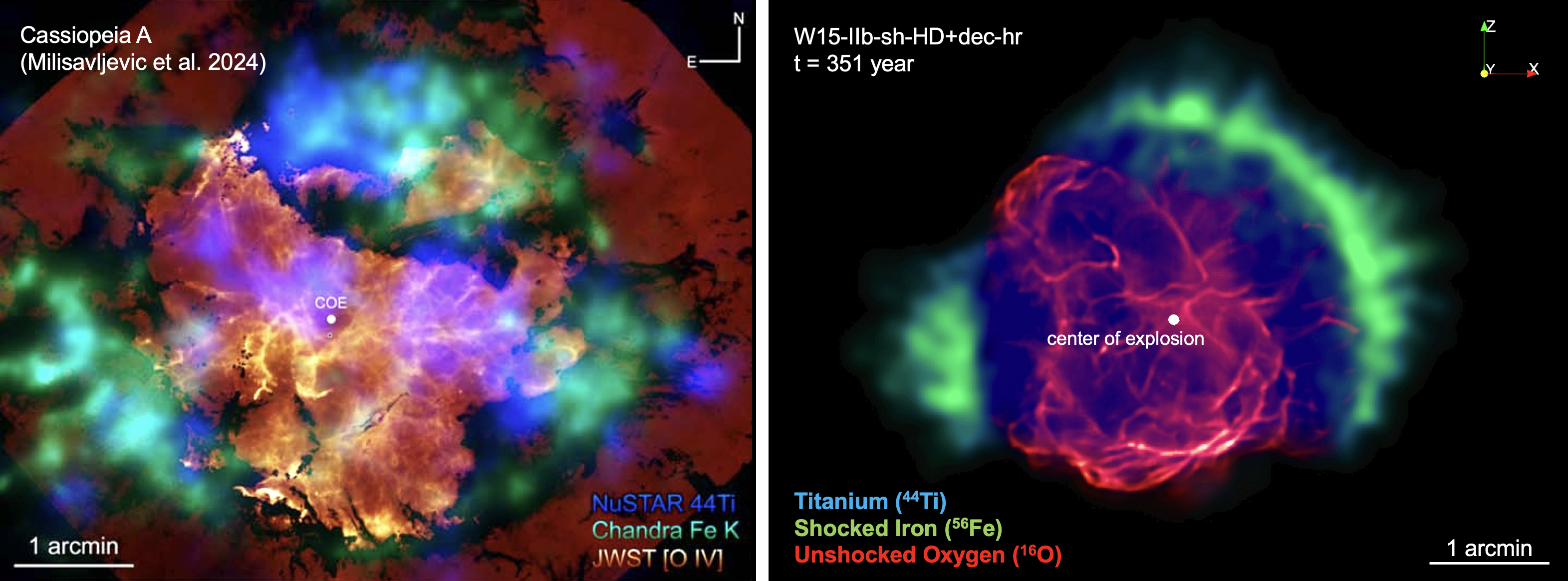}
   \caption{Left panel: Composite image (adapted from Fig.~6 in \citealt{2024ApJ...965L..27M}), combining the NuSTAR $^{44}$Ti map (blue), a Chandra Fe-K map (energy $\sim 6.7$~keV; green) with the
   unshocked ejecta map derived from JWST observations (red). The center of {\em expansion} (COE; \citealt{2001AJ....122..297T}) and the neutron star are indicated by white and black circles, respectively. Right panel: Composite image based on model W15-IIb-sh-HD+dec-hr (see Table~\ref{Tab:model}), showing 3D volumetric renderings of the distributions of $^{44}$Ti (blue), shocked $^{56}$Fe (green), and unshocked $^{16}$O (red). The center of {\em explosion} is marked with a white circle. While the model-derived image illustrates the overall geometry of the ejecta, it is not intended to replicate the exact observational appearance.}
   \label{whole_remnant}%
   \end{figure*}

First, we estimated the total mass of shocked and unshocked ejecta predicted by our models for \casa's current age, along with the mass fractions of the primary elements, to check how close to \casa\ our models are (see Table~\ref{Tab:elem}). We found that approximately $0.79\,M_{\odot}$ of ejecta have remained unshocked over 350 years. Although the SN model adopted (W15-2-cw-IIb; \citealt{2017ApJ...842...13W}) was not specifically fine-tuned to match \casa\ exactly, the estimated mass of unshocked ejecta is in reasonable agreement with values inferred from X-ray observations ($\approx 0.30\,M_{\odot}$; \citealt{2012ApJ...746..130H}), low-frequency radio observations ($<100$~MHz, $\approx 0.39\,M_{\odot}$; \citealt{2014ApJ...785....7D}), and recent infrared observations ($0.47^{+0.47}_{-0.24}\,M_{\odot}$; \citealt{2020ApJ...904..115L}), as well as with the value estimated from previous modeling studies ($\sim 0.34\,M_{\odot}$; \citealt{2016ApJ...822...22O}). In our model, most of the ejecta rich in helium (He) and carbon (C) have been shocked. The unshocked ejecta are predominantly composed of O (43\%) and He (14\%), with smaller fractions of C, calcium (Ca), magnesium (Mg), neon (Ne), and silicon (Si) (see Table~\ref{Tab:elem}). The iron (Fe) content, estimated at $0.064\,M_{\odot}$, is consistent with the upper limit derived from Spitzer data ($<0.07\,M_{\odot}$; \citealt{2020ApJ...904..115L}). Titanium (Ti) was found in minimal amounts, showing the lowest mass fractions among the elements considered. This is particularly notable given that the model overestimates Ti by a factor of $10-20$ due to the simplified reaction network employed in the original SN model (\citealt{2017ApJ...842...13W}). When this overestimation is accounted for, the Ti/Fe ratio derived from the model aligns closely with the ratio inferred from observations (e.g., \citealt{2021Natur.592..537S}).

Although discrepancies exist between the model and observations, particularly in element abundances, the comparison is satisfactory given the uncertainties in observational measurements and the use of a very small network in the original SN model, which led to an under- or overproduction of some species, for example the already mentioned overestimation of the abundance of $^{44}$Ti by a factor of $10-20$ (see \citealt{2017ApJ...842...13W}). Furthermore, our \casa\ model was not optimized for the explosion energy ($1.5 \times 10^{51}$~erg; \citealt{2017ApJ...842...13W}). A slightly higher explosion energy, closer to the observationally deduced value ($\sim 2\times 10^{51}$~erg; e.g., \citealt{2003ApJ...597..347L, 2020ApJ...893...49S}), could drive the reverse shock deeper into the ejecta, reducing the amount of unshocked material at the present epoch.

The left panel of Fig.~\ref{whole_remnant} presents a composite image adapted from Fig.~6 in \cite{2024ApJ...965L..27M}, combining NuSTAR, Chandra, and JWST observations to show the location of the O-rich filamentary network observed with JWST relative to the entire remnant. The right panel displays an analogous composite image derived from model W15-IIb-sh-HD+dec-hr, illustrating the distribution of unshocked O-rich ejecta (red) alongside shocked Fe-rich ejecta (green) and both shocked and unshocked Ti-rich ejecta (blue). To enable a meaningful comparison with the left panel, the Fe and Ti distributions in the model-derived image have been smoothed to approximate the spatial resolution of NuSTAR and Chandra observations shown in that panel.

It is worth noting that, while the model-derived image combines the distributions of O, Ti, and Fe to have an analog of the composite image on the left panel of the figure, it does not attempt to replicate the observational data. Achieving such a match would require a detailed synthesis of emissions in the specific bands observed by JWST, Chandra, and NuSTAR. Instead, the composite image from the model serves to illustrate the geometry of the three chemical components relative to each other and emphasize key morphological features. Indeed, a particularly striking feature in the model is the presence of a network of unshocked O-rich ejecta filaments, displaying general characteristics and location that are reminiscent of those observed with JWST. This similarity highlights the model's capability to explore the origin and evolution of this filamentary structure. Consequently, in the following we focused on examining the structure and dynamics of the unshocked ejecta, with a particular emphasis on the O-rich filaments and their physical origin.

   \begin{figure*}
   \centering
   \includegraphics[width=0.95\textwidth]{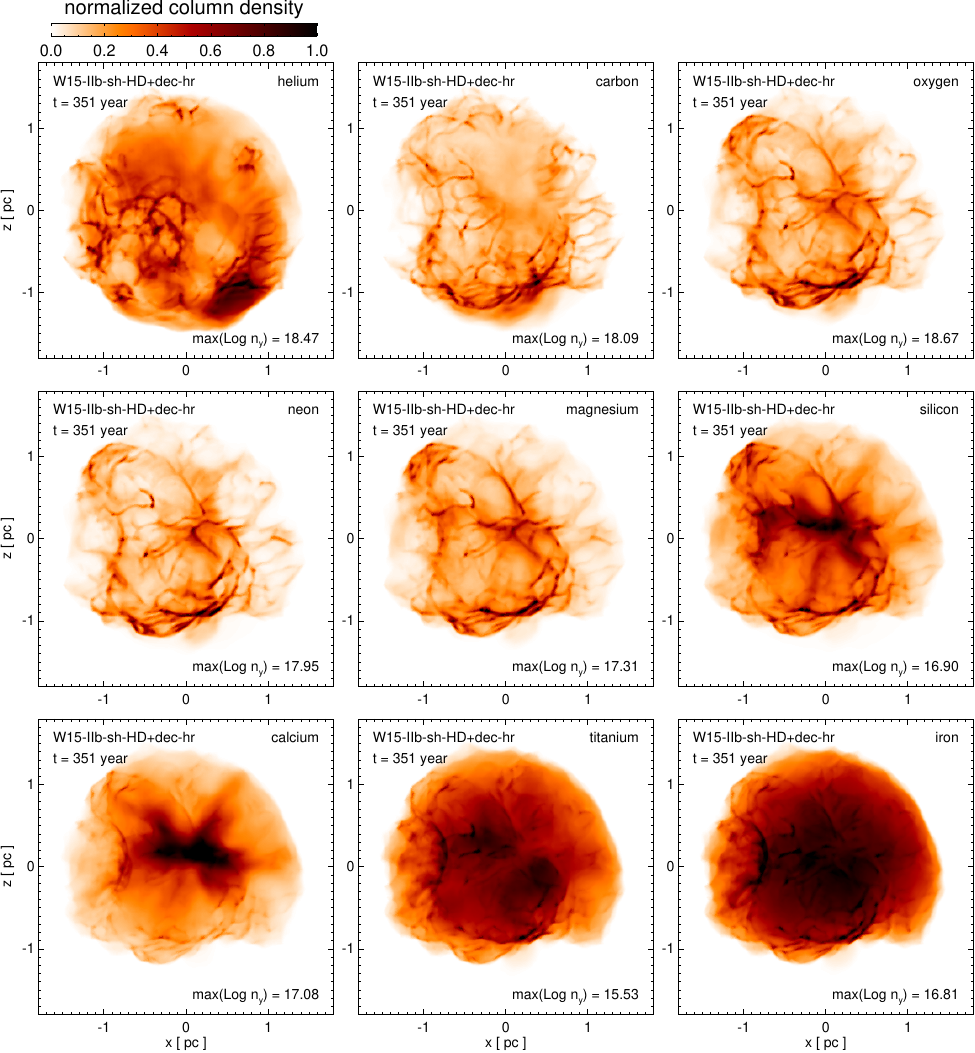}
   \caption{3D volume renderings of the mass density for unshocked ejecta enriched in the elements labeled in the upper right corner of each panel, integrated along the $y$-axis. The images are based on model W15-IIb-sh-HD+dec-hr at an age of 351 years, assuming the vantage point being at Earth, i.e. on the negative $y$-axis (the perspective is in the plane of the sky). Each image is normalized to its maximum value (shown in the lower-right corner of each panel in units of cm$^{-2}$) to enhance visibility. The color bar in the top-left corner of the figure indicates the normalized column density scale.}
   \label{maps_all_elem_front}%
   \end{figure*}

   \begin{figure*}
   \centering
   \includegraphics[width=0.95\textwidth]{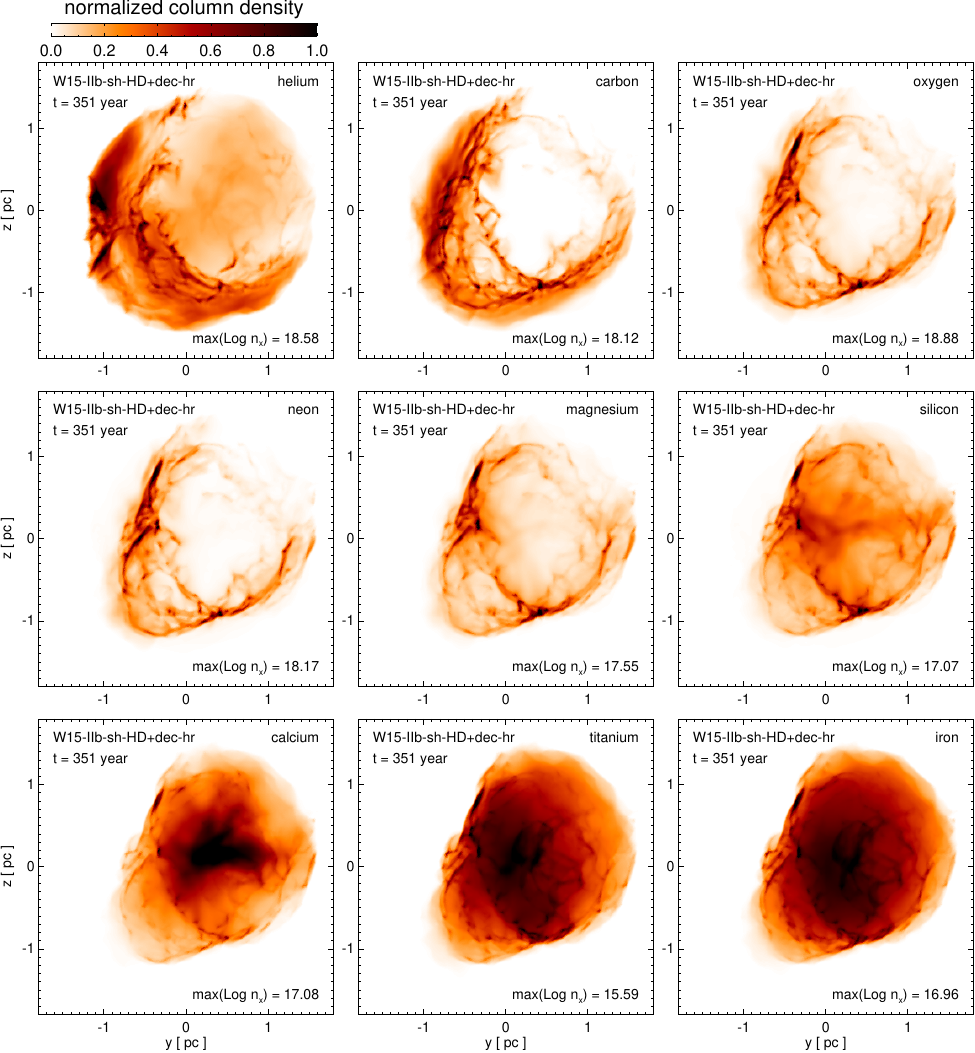}
   \caption{Same as Fig.~\ref{maps_all_elem_front}, but shown from an alternate viewing angle, with the vantage point positioned along the positive $x$-axis and the Earth being located along the negative $y$-axis. The images are integrated along the $x$-axis}
   \label{maps_all_elem_side}%
   \end{figure*}

   \begin{figure}
   \centering
   \includegraphics[width=0.47\textwidth]{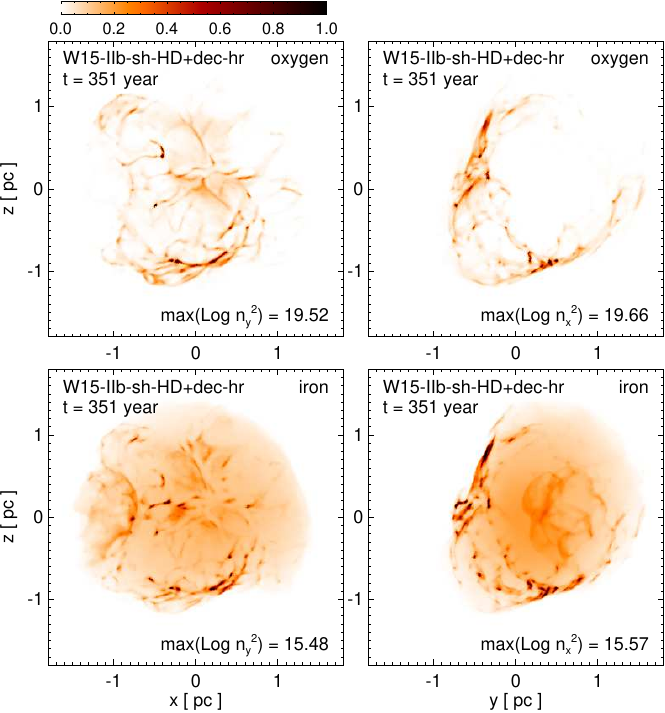}
   \caption{Same as Figs.~\ref{maps_all_elem_front} and \ref{maps_all_elem_side}, but shown for O (upper panels) and Fe (lower panels) considering 3D renderings of density squared instead of density. The images are integrated along either the $y$-axis (left panels) or the $x$-axis (right panel).}
   \label{maps_fe_o_n2}%
   \end{figure}

Figures~\ref{maps_all_elem_front} and \ref{maps_all_elem_side} present 3D volumetric renderings of the distributions of unshocked ejecta across different species, observed from two distinct viewing angles. These figures reveal that the intermediate-mass and light elements (He, C, O, Ne, Mg, Si) form an interconnected, web-like network of filaments. As anticipated by Fig.~\ref{whole_remnant}, this structure closely resembles the intricate, filamentary ejecta patterns observed by JWST in \casa, especially for O-rich ejecta (see Fig.~\ref{maps_all_elem_front}). In contrast, the filamentary network is notably less pronounced in the distributions of heavier elements such as Ca, Ti, and Fe. We note that normalizing the figures to the maximum value results in more saturated images for Fe compared to O. This suggests that Fe filaments are less locally dense, but the saturation also makes them harder to distinguish. To facilitate an easier comparison of the positions of Fe-rich filaments relative to O-rich ones, Fig.~\ref{maps_fe_o_n2} presents the 3D renderings of the density squared for O and Fe.

It is worth emphasizing that our simulations lack the fine-scale resolution necessary to replicate the smallest filaments detected by JWST. This limitation may be due to the spatial resolution constraints of the simulation: although the resolution is high, it may be not enough to prevent numerical diffusion from smoothing out the smallest-scale structures that are resolved by $< 10$ grid points. This limits the ability to capture features at a level comparable to JWST's observational capabilities. Despite this, our model effectively captures the broader filamentary structure, which is concentrated in a shell surrounding a central cavity filled with heavier elements (Ca, Ti, and Fe), including trace amounts of Si. This distribution is evident when comparing the lower panels of Fig.~\ref{maps_all_elem_side} with the upper and middle panels.

   \begin{figure*}
   \centering
   \includegraphics[width=\textwidth]{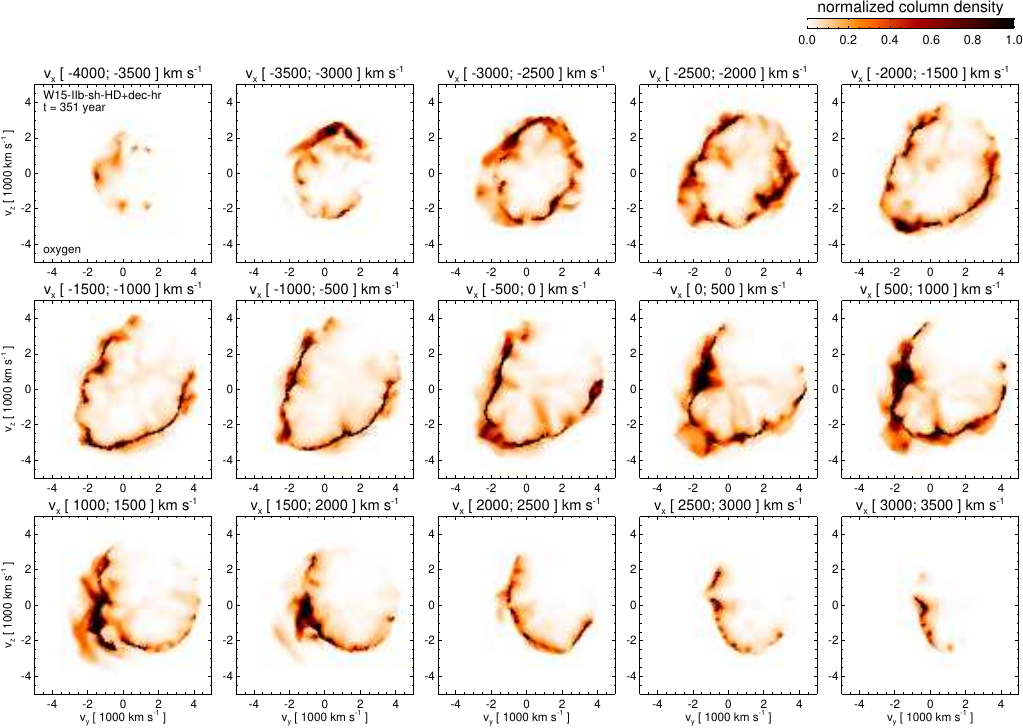}
   \caption{Distribution of O-rich ejecta as a function of velocity along the $x$-axis (from east to west in the plane of the sky), spanning from $v_{\rm x} = -4000$~km~s$^{-1}$ (top left panel) to $v_{\rm x} = 3500$~km~s$^{-1}$ (bottom right panel). Each frame represents the O-mass density integrated over a 500~km~s$^{-1}$ velocity interval along the $x$-axis (color bar on the top right of the figure), with the specific interval shown above each frame. Images are saturated at 10\% of the maximum value across all frames to enhance visibility. The axes represent the velocity components: $v_{\rm y}$ (LoS velocity, assuming the Earth is located along the negative $y$-axis) and $v_{\rm z}$ (perpendicular to the LoS, spanning south to north).}
   \label{velocity_distrib}%
   \end{figure*}

The filamentary network also reveals the presence of large voids, most prominently visible in the side view of the remnant (Fig.~\ref{maps_all_elem_side}), where intermediate-mass elements are notably scarce. These voids result from the extended Fe-rich plumes, which are relics of the Ni-rich plumes generated during the SN explosion (\citealt{2017ApJ...842...13W}). These plumes penetrated the outer layers of lighter-element ejecta and began interacting with the reverse shock as early as $\approx 30$ years after the explosion (\citealt{2021A&A...645A..66O}). This interaction has contributed to the formation of the prominent Fe-rich regions observed in \casa's main shell of shocked ejecta. This complex structure of the unshocked ejecta, with superposition of lighter filamentary networks with concentrations of heavier elements and Fe-rich plumes, is expected to reflect the large-scale mixing and distribution patterns emerging from the SN explosion.
   
Given that the filamentary network observed by JWST is rich in O, our analysis focused mainly on the simulated distribution of O-rich ejecta. Figure~\ref{velocity_distrib} presents this distribution as a function of the velocity of freely expanding material along the $x$-axis, ranging from east (negative velocities; top left in the figure) to west (positive velocities; bottom right) across the plane of the sky. The spatial scale is mapped onto a velocity scale, where $v_{\rm y}$ represents velocities along the line-of-sight (LoS; with negative values indicating blueshifts), and $v_{\rm z}$ represents velocities along the south-to-north direction. Each panel in Fig.~\ref{velocity_distrib} shows ejecta density integrated over 500~km~s$^{-1}$ intervals in $v_{\rm x}$, covering the range $v_{\rm x} = [-4000, 3500]$~km~s$^{-1}$. With Earth positioned along the negative $y$-axis, this representation effectively highlights the structure and distribution of O-rich ejecta along the LoS, providing valuable insights into the spatial dynamics of the filamentary network.

The O-rich ejecta in Fig.~\ref{velocity_distrib} are concentrated within a thin shell with LoS velocities, $v_{\rm y}$, ranging from $-3000$ to $+4500$~km~s$^{-1}$, consistent with the velocity distribution of unshocked ejecta inferred from the analysis of \casa\ observations (\citealt{2015Sci...347..526M,2024ApJ...965L..27M}). This shell opens toward the northwest on the redshifted side of the remnant, with a significant absence of O-rich material indicated by the incomplete rings in the panels of Fig.~\ref{velocity_distrib} at velocities $v_{\rm x} > -1500$~km~s$^{-1}$. This depletion is primarily due to Fe-rich ejecta embedded within the shell, which extend outward into the northern redshifted region of the remnant (see Fig.~7 in \citealt{2021A&A...645A..66O}), disrupting the continuity of the O-rich shell. This Fe-rich plume has been interacting with the reverse shock since approximately 30 years post-explosion.

The highest concentration of O-rich ejecta occurs in a blueshifted region, with LoS velocities around $v_{\rm y} \approx -2000$~km~s$^{-1}$. This region aligns with a bundle of filaments forming an extended arc-like structure visible near the remnant's center in Fig.~\ref{maps_all_elem_front} (top right). A similar bundle of filaments is observed in the distributions of Ne, Mg, and Si, as also indicated by the analysis of Spitzer data on \casa\ (\citealt{2010ApJ...725.2059I}). These features are consistent with JWST observations of the web-like network in \casa. However, the velocities of the O-rich ejecta inferred from the observations exhibit a nearly continuous range (\citealt{2024ApJ...965L..27M}; Dickinson et al., in preparation), suggesting the presence of O-rich filaments in the inner ejecta region.

\begin{figure*}
   \centering
   \includegraphics[width=\textwidth]{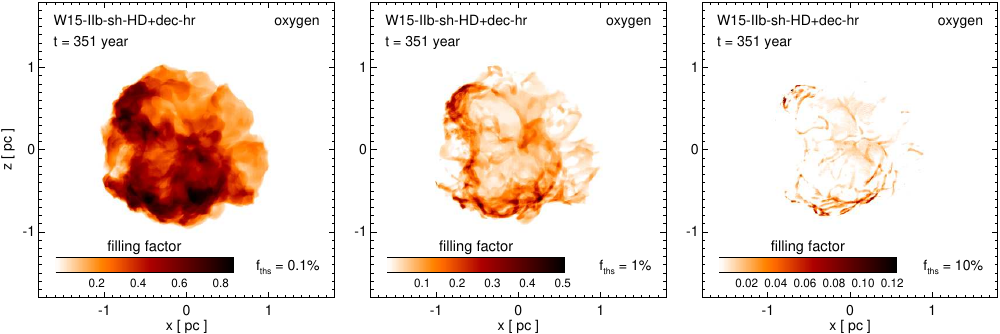}
   \caption{Maps of the filling factor of O-rich ejecta, calculated for cells where the O density exceeds thresholds of 0.1\% (upper panel), 1\% (middle panel), or 10\% (bottom panel) of the maximum O-rich ejecta density in the unshocked volume. The maps assume the model orientation matching \casa.}
   \label{filling}%
\end{figure*}

\begin{figure}
   \centering
   \includegraphics[width=0.45\textwidth]{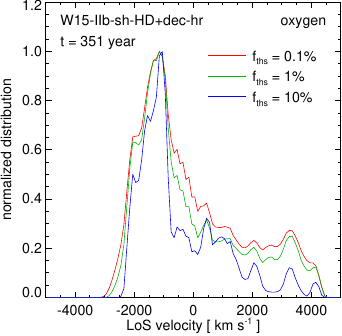}
   \caption{Normalized LoS velocity distributions corresponding to the maps of the filling factor of O-rich ejecta shown in Fig.~\ref{filling}, derived from the same cells used to derive the filling factor maps.}
   \label{filling_vel}%
\end{figure}

To investigate further this issue and to facilitate a more quantitative comparison between model predictions and JWST observations, we calculated the filling factor of O-rich ejecta and the corresponding LoS velocity distributions. Figure~\ref{filling} displays maps of the filling factor, assuming the model is oriented to match \casa. These maps represent the filling factors computed along the LoS for cells where the O density exceeds specific thresholds ($f_{\rm ths} =$ 0.1\%, 1\%, or 10\%) of the maximum O-rich ejecta density within the entire unshocked volume. Specifically, for each LoS, the filling factor is calculated as the ratio of the volume of cells containing O-rich ejecta above a given density threshold ($V_{\rm O}$) to the total volume of all cells containing unshocked ejecta along the same LoS ($V_{\rm tot}$). This ratio provides a normalized measure of the extent to which the unshocked ejecta volume is occupied by O-rich material, offering valuable insights into its spatial distribution and density contrasts. Figure~\ref{filling_vel} shows the corresponding normalized LoS velocity distributions. These distributions are derived from the same cells used to compute the filling factors and highlight the kinematic properties of the O-rich ejecta, complementing the spatial analysis provided by the maps.

According to Fig.~\ref{filling}, although the O density is significantly lower than in the prominent dense shell shown in Figs.~\ref{maps_all_elem_front} and \ref{maps_all_elem_side}, O-rich ejecta are also present in the innermost regions of the remnant, which are rich in Fe and other heavy elements (as evident from Fig.~\ref{maps_fe_o_n2}). Lowering the density threshold for selecting cells substantially increases the filling factor, reaching approximately 0.5 at a 1\% threshold and about 0.8 at a 0.1\% threshold of the maximum O-rich ejecta density (see upper and middle panels in the figure). Although web-like structures and filaments rich in O and Fe can also be seen in the innermost ejecta  (Figs.~\ref{maps_fe_o_n2} and \ref{filling}), these features may appear a bit less sharp and less concentrated than in the outer regions. This visual effect may be connected to differences in the relative spatial resolution $\delta r/r$ when a uniform cartesian grid with cell size $\delta x$ is applied to resolve structures at different radii $r$. In Appendix \ref{app:resolution}, we evaluated the impact of spatial resolution on structures forming at different distances $r$ from the explosion center. It may be possible that filaments forming in the innermost regions of the remnant are resolved with lower accuracy, leading to slightly smoother and more diffuse structures, as observed in the simulations.

The presence of O-rich ejecta in the remnant's interior results in a LoS velocity distribution characterized by a prominent peak at $u_{\rm LoS} \approx -1300$~km~s$^{-1}$ (see Fig.~\ref{filling_vel}), consistent with expectations from Fig.~\ref{velocity_distrib}. Moreover, the velocity distribution extends almost continuously up to $u_{\rm LoS} \approx 4500$~km~s$^{-1}$, exhibiting a broad range of velocities between $-2500$ and $4500$~km~s$^{-1}$. We also observe a pronounced asymmetry in the velocity distribution of unshocked ejecta, with material moving toward us reaching maximum absolute velocities of $\approx -2500$~km~s$^{-1}$, while ejecta traveling away exhibit maximum absolute velocities of about $\approx +4500$~km~s$^{-1}$. This result is consistent with the analysis of Spitzer observations of \casa\ (\citealt{2010ApJ...725.2059I}). The extended asymmetric velocity distribution underscores the complex kinematics of the O-rich ejecta and highlights their dynamic mixing within the remnant's interior.

\begin{figure*}
   \centering
   \includegraphics[width=\textwidth]{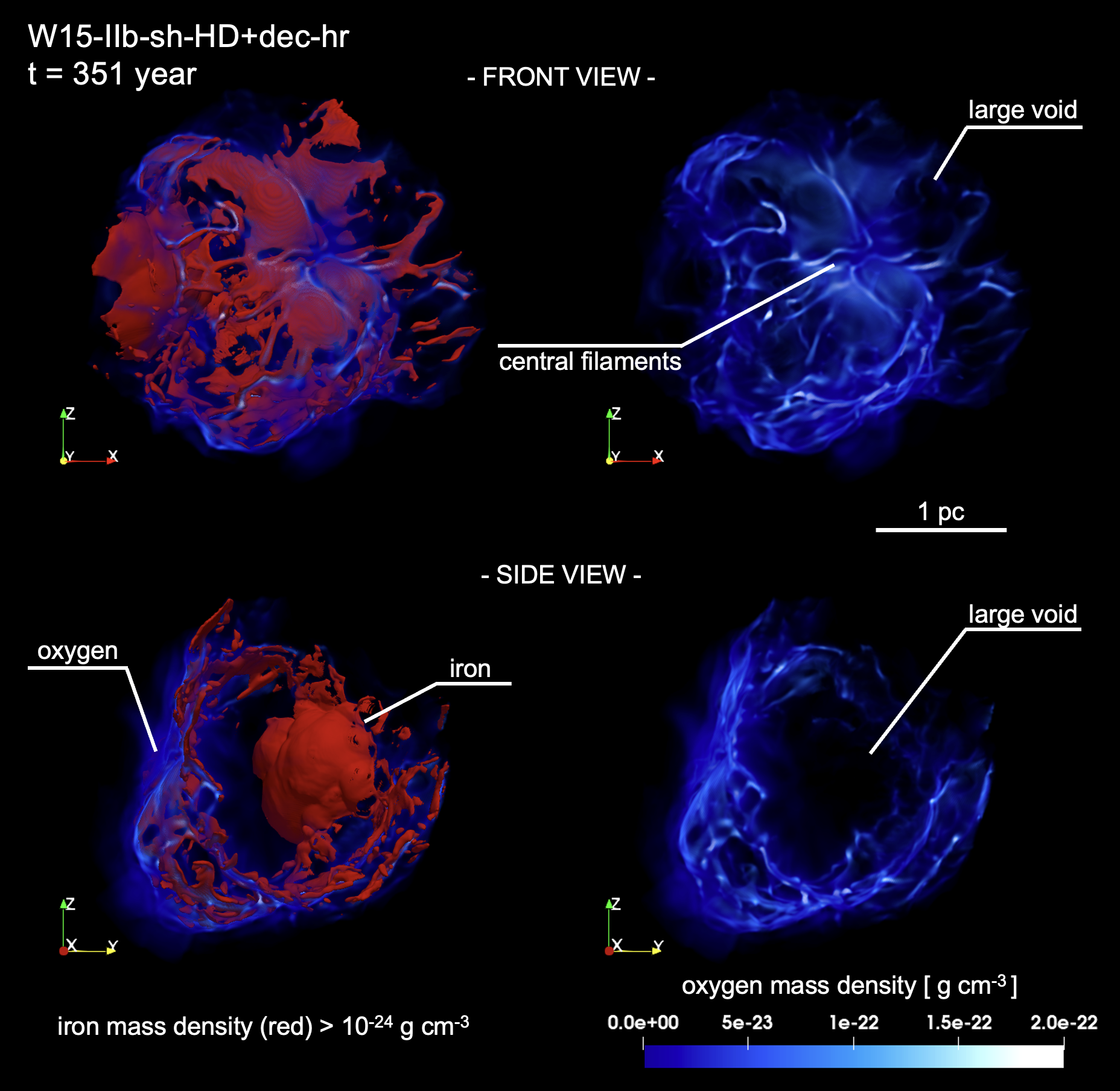}
   \caption{Distribution of unshocked Fe-rich ejecta, represented as a red isosurface in the left panels, corresponding to Fe densities exceeding $10^{-24}$~g~cm$^{-3}$, at the age of \casa\ for model W15-IIb-sh-HD+dec-hr. The unshocked O-rich ejecta are shown through volume rendering in a blue color palette, with the color scale provided at the bottom right, both in combination with the Fe-rich distribution (left panels) and independently (right panels). The opacity in the rendering is proportional to the plasma density, enhancing the visualization of denser regions. Two perspectives are presented: the upper panels show the front view as seen from Earth, while the lower panels provide a side view from a vantage point to the west (positive $x$-axis). See online Movie 1 for an animation of these data; a navigable 3D graphic of the O and Fe spatial distributions at the age of \casa\ is available at {https://skfb.ly/psXKr}.}
   \label{ejecta_structure}%
   \end{figure*}

The derived filling factors and LoS velocity distributions enable a preliminary comparison with observed spatial distributions, offering insights into how well the model reproduces the observed 3D structure of O-rich ejecta. However, some words of caution are needed in interpreting these comparisons. The filling factor maps from the simulation do not directly correspond to those of observational data, as the entire unshocked volume cannot be probed in observations as we do in the model. Instead, observations are limited to pencil or box beams (e.g., at positions P2 and P4 in Fig.~4 of \citealt{2024ApJ...965L..27M}) that sample the unshocked inner volume. Within these beam volumes, only the local contrast in O densities can be detected, inferred from line intensities, rather than the global distribution. Despite these limitations, comparing the simulation-derived maps in Fig.~\ref{filling} with observational data provides valuable context for evaluating the model's fidelity in capturing the morphology and dynamics of the ejecta, as well as for identifying potential discrepancies between the model and observational data. This comparison can reveal critical aspects of the physical processes shaping the remnant, helping to refine both the model and our understanding of the system.

   \begin{figure*}
   \centering
   \includegraphics[width=0.95\textwidth]{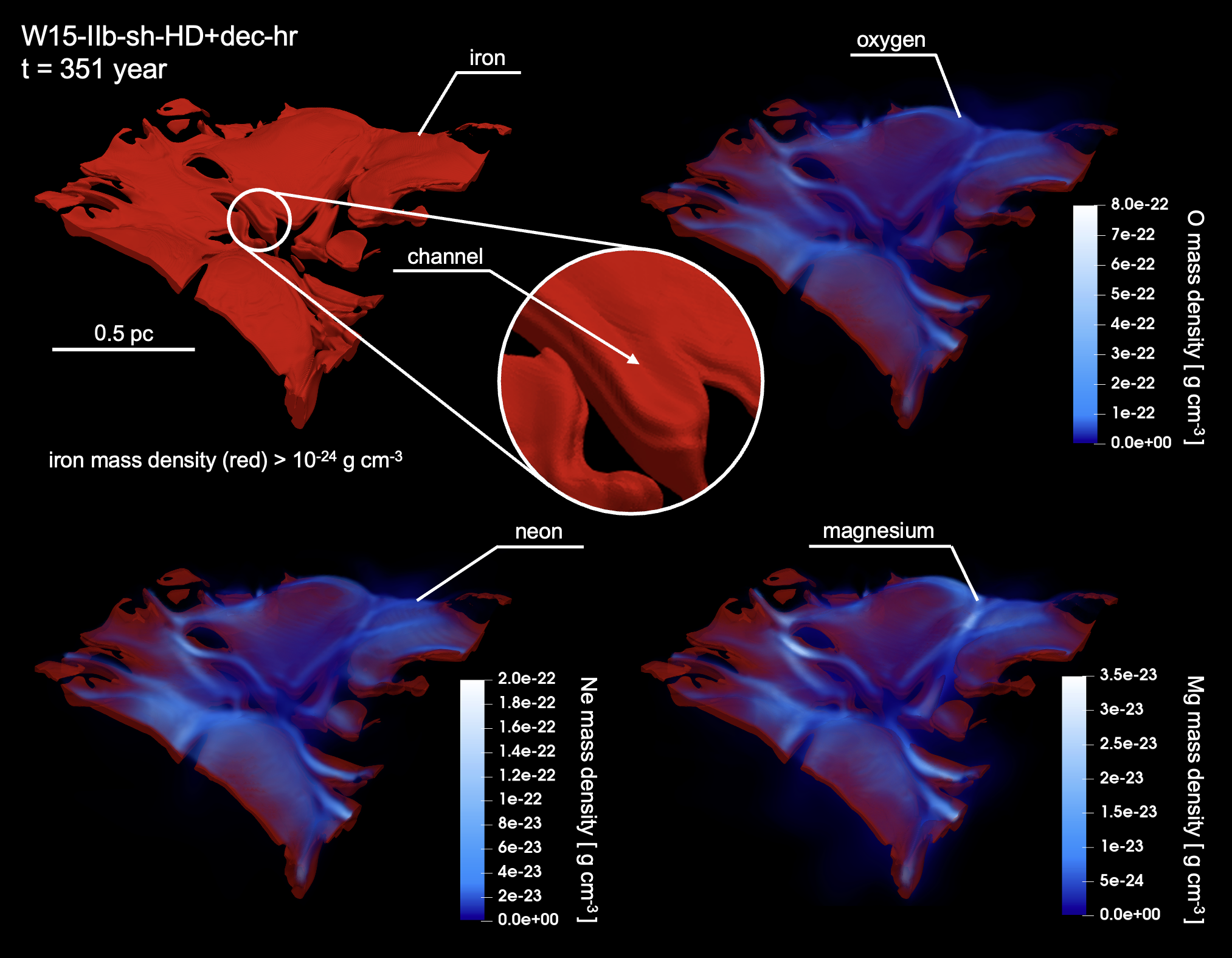}
   \caption{Close-up view of the region containing the central bundle of filaments highlighted in Fig.~\ref{ejecta_structure}. The unshocked Fe-rich ejecta with densities greater than $10^{-24}$~g~cm$^{-3}$ are represented as a red isosurface. The corresponding distributions of ejecta enriched in O (top right), Ne (bottom left), and Mg (bottom right) are visualized through volume rendering using a blue color palette. Each panel includes a color scale on the right, indicating the density  variations of the respective elements.}
   \label{channels}%
   \end{figure*}
   
Since the O-rich network of filaments includes an area of Fe-rich ejecta, we investigated potential spatial correlations between these structures. Figure~\ref{ejecta_structure} shows the distribution of the densest component of unshocked Fe-rich ejecta (red isosurface), where the mass density is $\rho_{\rm Fe} > 10^{-24}$~g~cm$^{-3}$. When overlaid with the O-rich filaments, an interesting spatial correlation emerges (see upper left panel in the figure and online Movie 1): the filaments systematically align with depressions or sort of channels in the surface of high-density Fe-rich ejecta. This alignment is particularly evident in the structure characterized by a bundle of filaments at the center of the remnant, where the filaments appear to trace the topographical contours of the Fe ejecta (see also online Movie 1). 

To investigate this correlation more deeply and examine the distribution of additional chemical species, we focused our analysis on the central bundle of filaments with the aim to explore the underlying physical mechanisms responsible for the observed filamentary network. Figure~\ref{channels} provides a close-up view of the region encompassing the central bundle of filaments, previously highlighted in Fig.~\ref{ejecta_structure}. The upper left panel displays the distribution of Fe-rich ejecta, represented by a red isosurface corresponding to densities $> 10^{-24}$~g~cm$^{-3}$. The orientation assumes the explosion center is positioned at the bottom, with the ejecta expanding upward. The surface thus delineates the region dominated by Fe-rich ejecta and exhibits a highly irregular morphology, featuring numerous depressions and holes. The cavities are filled with Fe-rich ejecta at lower densities, creating a complex and uneven topology.

   \begin{figure*}
   \centering
   \includegraphics[width=\textwidth]{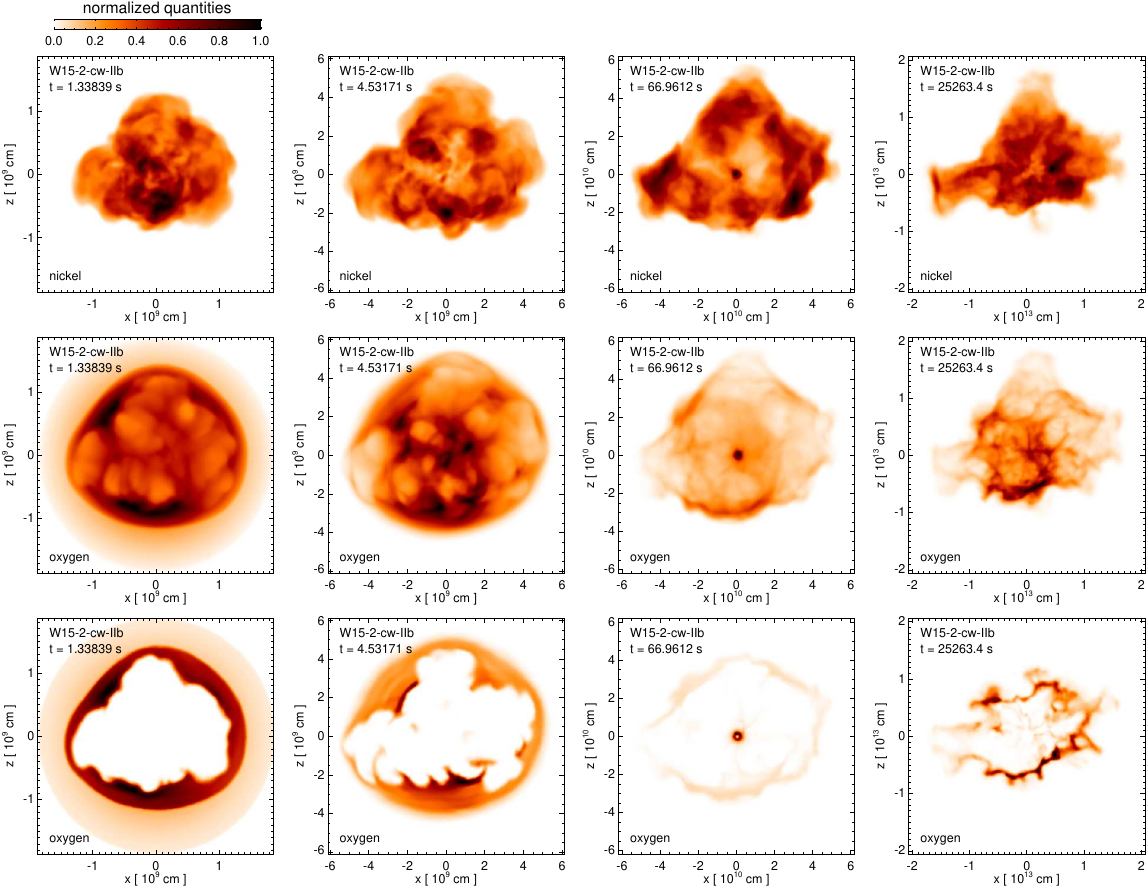}
   \caption{Spatial distributions of Ni and O during the first 7 hours of evolution following core collapse, based on model W15-2-cw-IIb (\citealt{2017ApJ...842...13W}). The upper and middle panels present volumetric renderings of Fe- and O-rich material distributions, respectively, at the specified times. The lower panels display 2D cross-sections of the O-rich ejecta in the $[x,z]$ plane, intersecting the explosion center. Each image is normalized to each maximum for visibility (color bar on the top left of the figure). See online Movie 2 for an animation of these data.}
   \label{sn_evolution}%
   \end{figure*}
   
We note that the isosurface is characterized by distinct channels (an example is shown in Fig.~\ref{channels}), constrained by high density Fe-rich ejecta. These channels are occupied by ejecta rich in O (top right panel), Ne (bottom left panel), and Mg (bottom right panel), visualized through volume rendering with a blue color palette. The opacity in these renderings is proportional to plasma density, emphasizing regions of higher concentrations. Similarly to rivers filling valleys, O-, Ne-, and Mg-rich ejecta pervades the channels or grooves in the Fe-rich structure forming the intricate web-like pattern of ejecta observed by JWST. We found that the filaments exhibit thicknesses down to scales of $\approx 0.01$~pc, consistent with the dimensions of the observed filaments (\citealt{2024ApJ...965L..27M}, Dickinson et al., in preparation).
   
This complex spatial arrangement suggests a physical interaction between the various ejecta layers during the explosion and subsequent expansion. These interactions are likely driven by processes such as HD instabilities, mixing, or differential velocities between matter enriched with different chemical species. A notable example are the Fe-rich ejecta plumes that contribute to the inversion of ejecta layers observed in \casa\ (see \citealt{2016ApJ...822...22O, 2021A&A...645A..66O} for a detailed discussion).

\subsection{Generation of the web-like network of ejecta filaments}
\label{sec:res2}

The filamentary structures reproduced in our model resemble those observed by JWST in \casa\ and exhibit comparable properties (e.g., \citealt{2024ApJ...965L..27M}). We note that all the features shaping the ejecta structure in our model emerge from stochastic processes that naturally occur in the aftermath of the core collapse (\citealt{2017ApJ...842...13W}), without relying on ad-hoc assumptions. By tracing the evolution of these features from the first seconds after the core-collapse to the current condition, therefore, we can gain valuable insights into the mechanisms responsible for shaping the observed filamentary network. In this way, we were able to identify the physical processes driving the complex phases of the SN evolution, for which the filaments are the observable signatures. The complete evolution from $\approx 1.3$~s after the core collapse to the age of $\approx 1000$~yr is provided as online Movie 2.

According to the SN model adopted in our study (W15-2-cw-IIb; \citealt{2017ApJ...842...13W}), the first few seconds following the core collapse witness the emergence of expanding bubbles of neutrino-heated material growing in the dense postshock plasma around the new-born neutron star. These buoyant bubbles rapidly ascend, profoundly altering the star's original chemical stratification (see Fig.~\ref{sn_evolution}, panels in the first two columns on the left; see also Movie 2). As they rise, they penetrate and destabilize the overlying stellar layers, generating extensive regions of turbulent mixing. This process lifts heavier elements, such as nickel (Ni; upper panels of the figure), into regions previously dominated by lighter materials, such as O (lower panels). The resulting turbulent interactions create intricate, non-uniform structures that deviate significantly from the star's initial stratified, onion-shell-like configuration.

   \begin{figure*}
   \centering
   \includegraphics[width=0.97\textwidth]{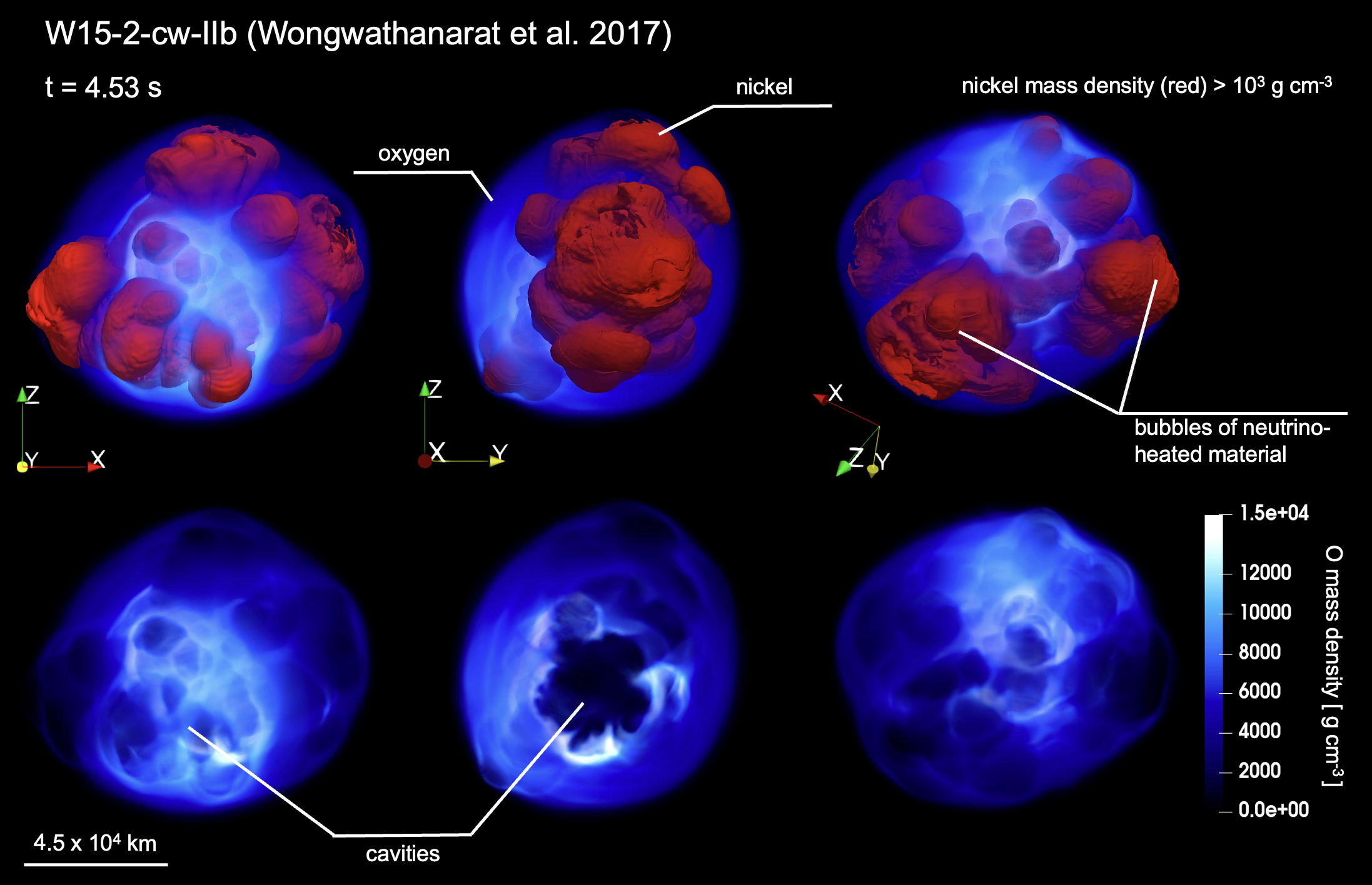}
   \caption{The upper panels display red isosurfaces representing the distribution of Ni-rich ejecta (at a density threshold of $10^3$~g~cm$^{-3}$) at $t = 4.53$~s after core collapse for model W15-2-cw-IIb (\citealt{2017ApJ...842...13W}), shown from different viewing angles. The corresponding O-rich ejecta are visualized using a blue volume rendering (with opacity scaled to reflect density variations), both in combination with the Ni-rich distribution (upper panels) and independently (lower panels). The bottom right color bar indicates plasma density.}
   \label{bubbles}%
   \end{figure*}

As the network of expanding bubble structures propagates outward, it initially carves out large cavities in the distribution of the overlying stellar material (see lower panels in Fig.~\ref{sn_evolution} for $t < 5$~s, and Movie 2), filling them with neutrino-heated material (upper panels in the figure). For instance, Fig.~\ref{bubbles} illustrates the distributions of Ni- and O-rich ejecta approximately 4~s after the core collapse, viewed from different angles. At this stage, bubbles of neutrino-heated material rich in Ni (red isosurface) have penetrated the overlying O-rich layer (blue volumetric rendering). In regions where the fast-rising Ni-rich plumes compress the O-rich material, dense sheaths of O-rich ejecta are formed, eventually enveloping portions of the Ni-rich plumes. These structures are particularly evident in the lower panels of Fig.~\ref{bubbles}. Furthermore, the intricate topology of the expanding bubble network often causes dense sheets of O-rich ejecta to become trapped at the boundaries where two or more bubbles intersect. These processes play a crucial role in shaping the distribution and mixing of elements within the ejecta, leading to the development of highly non-uniform structures.

During this phase of evolution, the neutrino-heated bubbles rapidly grow, expanding from initial structures on the scale of tens of kilometers (shortly after the core collapse) to vast regions spanning thousands of kilometers within just a few seconds (see second column from the left in Fig.~\ref{sn_evolution}, and Movie 2). In the subsequent phases, the morphology of these bubbles evolves dramatically, transitioning from smooth, nearly spherical shapes to highly complex, cellular, and interconnected volumes with intricate, irregular boundaries (see Fig.~\ref{sn_evolution}, panels in the last two columns on the right, and Movie 2). As a result of this dynamics, the original O-rich layer of stellar progenitor becomes heavily modified, with portions being compressed, streatched, or swept away entirely. The dense sheaths of O-rich material that form around the ascending Ni-rich plumes are subject to further stretching and compression due to the plumes' continued growth and interaction with the surrounding medium. This process gives rise to a primordial network of dense sheaths and filamentary structures, forming an intricate web that becomes progressively more pronounced as the system evolves. By approximately 7 hours after the core collapse, these complex structures are already distinctly visible (see the right panels of Fig.~\ref{sn_evolution}, Movie 2, and for explosions of red and blue supergiant progenitors also discussed in \citealt{2015A&A...577A..48W}).

   \begin{figure*}
   \centering
   \includegraphics[width=\textwidth]{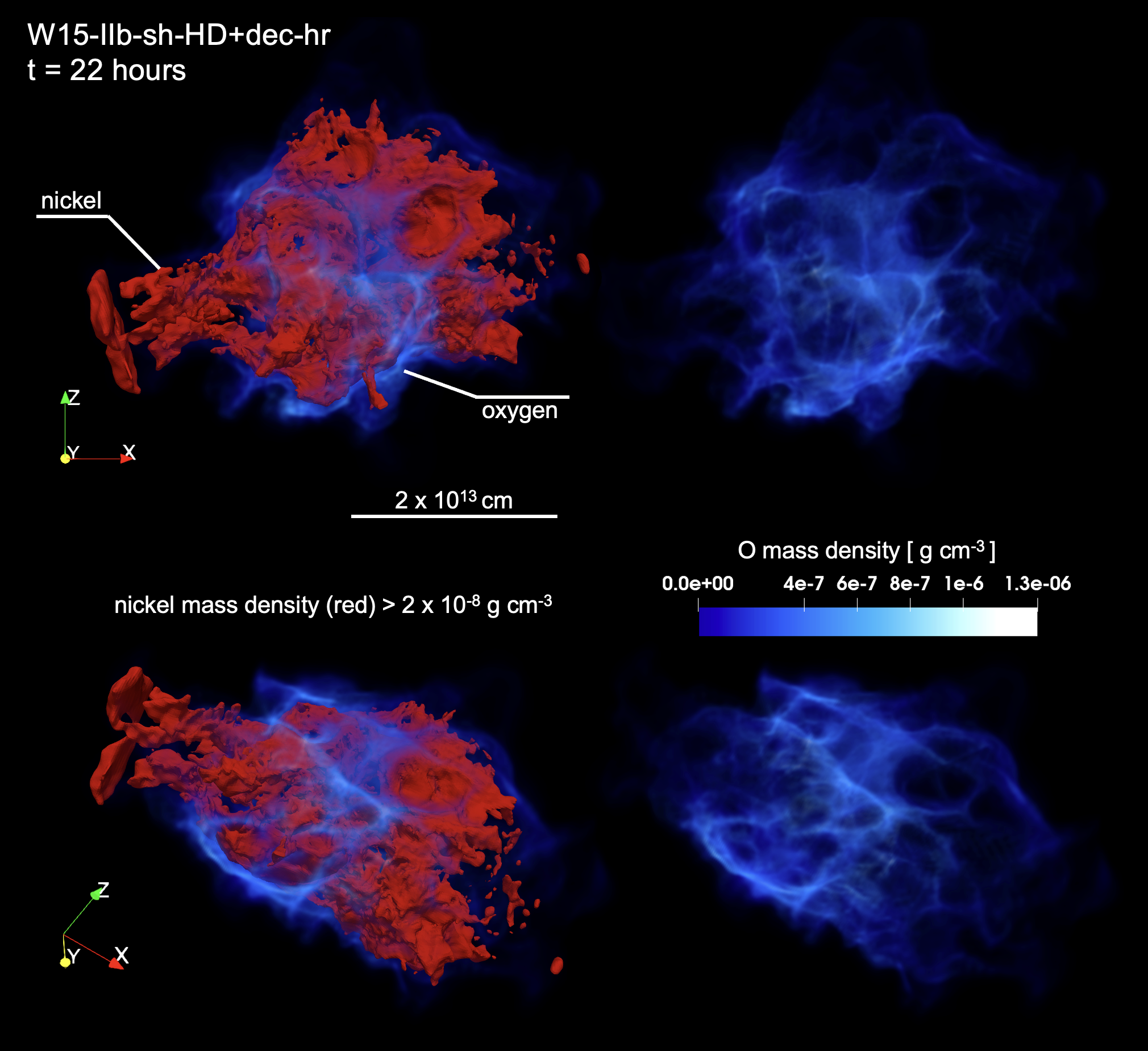}
   \caption{Distribution of unshocked Ni-rich ejecta represented as a red isosurface in the left panels, corresponding to Ni densities exceeding $2\times 10^{-8}$~g~cm$^{-3}$, a few hours after the shock breakout $\sim 22$~hours after the core-collapse, for model W15-IIb-sh-HD+dec-hr. The unshocked O-rich ejecta are shown through volume rendering in a blue color palette, as in Fig.~\ref{ejecta_structure}. Two perspectives are presented: the upper panels show the front view as seen from Earth, while the lower panels provide the perspective from an arbitrary point of view. A navigable 3D graphic of the O and Ni spatial distribution in this phase is available at {https://skfb.ly/psXKs}.}
   \label{sh_breakout}%
   \end{figure*}

Figure~\ref{sh_breakout} shows the distributions of Ni- and O-rich ejecta roughly one day after the shock breakout, approximately 22~hours after the core collapse. The densest regions of Ni-rich ejecta, with densities exceeding $2 \times 10^{-8}$~g~cm$^{-3}$, are displayed as a red isosurface in the left panels. At this stage, the web-like network of O-rich ejecta filaments is already well developed (see left panels), showcasing its intricate complexity. According to the evolution before the shock breakout described above, this structure reflects the turbulent mixing induced by the expansion of bubbles of neutrino-heated material and the effects of HD instabilities (see also \citealt{2015A&A...577A..48W, 2024arXiv241103434V}). In other words, the network of filaments underscores the dynamic interplay of neutrino-driven convection, which accelerates the efficient mixing of heavy elements as the blast wave propagates outward through the progenitor star's layers. Additionally, the figure reveals notable inversions of the ejecta layers in specific regions, consistent with observations of similar phenomena in \casa, as discussed in previous studies (\citealt{2016ApJ...822...22O, 2021A&A...645A..66O}). For instance, in the extended Ni-rich plume expanding eastward, O-rich ejecta are found to occupy a physically interior position relative to Ni-rich material. Such layer inversions are indicative of the highly asymmetric and turbulent nature of the explosion, driven by the initial anisotropies in the neutrino heating and further amplified during the expansion.

   \begin{figure*}
   \centering
   \includegraphics[width=\textwidth]{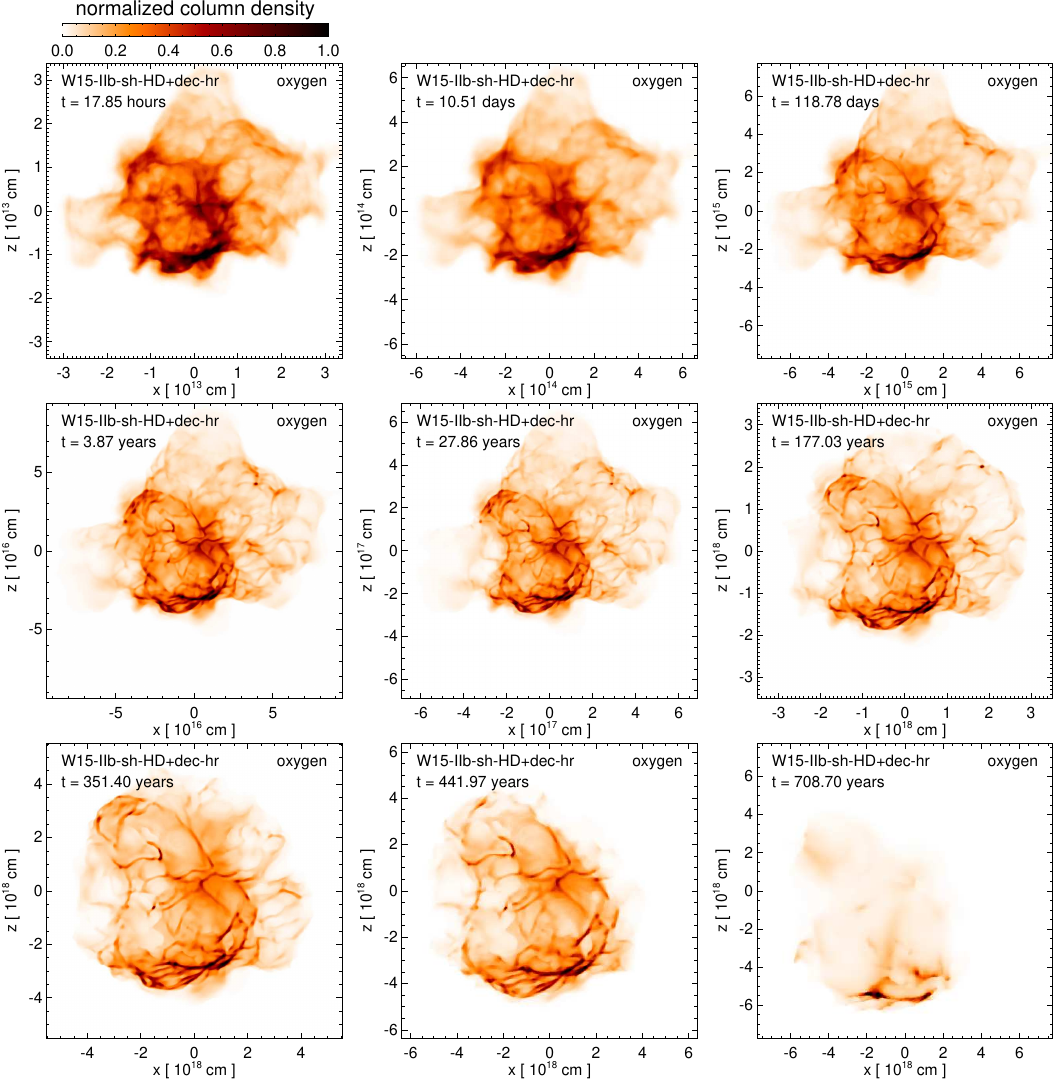}
   \caption{Volume renderings of the mass density distribution for unshocked O-rich ejecta, integrated along the LoS. The images are based on model W15-IIb-sh-HD+dec-hr at the labeled times (indicated in the upper left corner of each panel) and each of them is normalized to its maximum (color bar on the top left of the figure). The perspective assumes a vantage point from Earth, located along the negative y-axis, corresponding to the plane of the sky. See online Movie 2 for an animation of this data.}
   \label{evolution}%
   \end{figure*}

In the subsequent phase of evolution, the ejecta expand through the CSM, possibly interacting with local inhomogeneities of the ambient environment (see Fig.~\ref{evolution} and Movie 2). In our model, the main inhomogeneity encountered by the remnant is the dense asymmetric shell located at a distance of $\approx 1.5$~pc from the center of explosion (\citealt{2022A&A...666A...2O}). During the first year, the expansion of the innermost ejecta is not homologous due to the effects of energy deposition from the dominant radioactive decay chain $^{56}$Ni $\rightarrow$ $^{56}$Co $\rightarrow$ $^{56}$Fe (see also \citealt{2021MNRAS.502.3264G}). The localized heating induced by this process increases the pressure within the Ni-rich regions, causing them to expand more rapidly than their surroundings. The resulting overpressure creates "bubbles" of hot, low-density material that push outward against the denser, cooler ejecta. These structures are known as Ni-bubbles.

   \begin{figure*}
   \centering
   \includegraphics[width=0.93\textwidth]{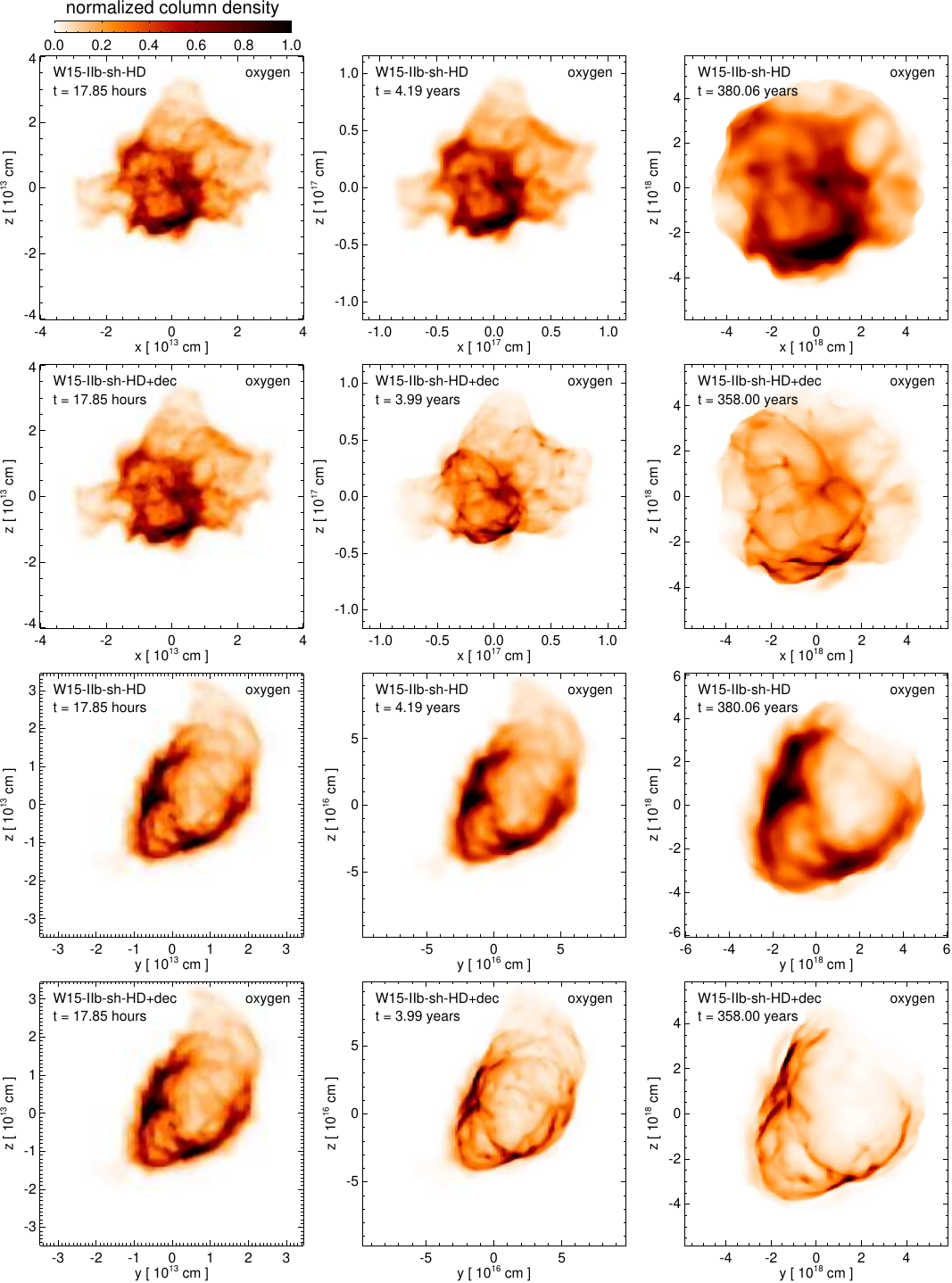}
   \caption{Comparison of simulations either with (W15-IIb-sh-HD+dec) or without (W15-IIb-sh-HD) the Ni-bubble effect. The figure presents volumetric renderings of the mass density distribution for unshocked O-rich ejecta, integrated along the LoS, at the labeled times. Each image is normalized to its maximum for better visibility (color bar on the top left of the figure). The first two rows show the simulations from the perspective of Earth, corresponding to the negative $y$-axis (front view). The last two rows provide the corresponding volume renderings from a vantage point along the positive $x$-axis (side view). }
   \label{compare_ni_bubble}%
   \end{figure*}

Our model underscores the pivotal role of the Ni-bubble effect in the final shaping of the web-like network of ejecta filaments, yielding a morphology closely resembling that observed in \casa. Figure~\ref{compare_ni_bubble} compares the distribution of O-rich ejecta obtained from our intermediate-resolution simulations (runs W15-IIb-sh-HD and W15-IIb-sh-HD+dec, described in Table~\ref{Tab:model}), either with or without the Ni-bubble effect included. Although the spatial resolution of these simulations does not fully resolve the small-scale features evident in our high-resolution runs (see Appendix~\ref{app:resolution}), the main filaments forming the network are captured in both cases. Notably, the filaments in run W15-IIb-sh-HD+dec, which includes the Ni-bubble effect, appear thinner, denser, and brighter compared to those in W15-IIb-sh-HD. This difference arises from the localized heating induced by the decay of radioactive $^{56}$Ni and its product, $^{56}$Co, which drives an enhanced expansion of the originally Ni-rich material. Given the half-lives of $^{56}$Ni ($\sim 6$~days) and $^{56}$Co ($\sim 77$~days), we expect the Ni-bubble effect to be significant during the first few months after shock breakout.

The enhanced expansion compresses and confines the O-rich, as well as C-, Ne-, Mg-, and Si-rich filaments trapped along the boundaries of the original Ni-rich regions, resulting in sharper and more compact structures. In contrast, in the absence of the Ni-bubble effect, the ejecta expand homologously only a few hours after the shock breakout (see upper panels in Fig.~\ref{compare_ni_bubble}); the ejecta filaments experience less localized pressure and are consequently broader and less defined. This result, therefore, highlights the role of the Ni-bubble effect in shaping the filamentary structure of unshocked ejecta to their final morphology, contributing to the complexity and brightness observed in \casa.

After one year of evolution, the remnant continues its expansion into the CSM while a reverse shock propagates backward through the ejecta. Once the Ni-bubble phase concludes, the innermost ejecta begin to expand almost homologously, preserving their structure (see Movie 2). This homologous expansion allows the innermost ejecta to keep a "memory" of the physical processes that occurred before and shortly after the shock breakout, which are responsible for their intricate filamentary morphology. 

Approximately 30~yr after the explosion, the Fe-rich ejecta start to interact with the reverse shock. This marks the onset of the processes that form the Fe-rich regions observed in \casa\ (see \citealt{2021A&A...645A..66O} for a detailed discussion). During this interaction, the reverse shock compresses and heats the Fe-rich ejecta, amplifying asymmetries in their distribution. The model also highlights the significant impact of the reverse shock on the ejecta filaments, particularly through the development of HD instabilities (primarily Rayleigh-Taylor and Kelvin-Helmholtz modes) at the contact discontinuity between the shocked ejecta and the shocked CSM. These instabilities, strongly perturb the filamentary structures, leading to their progressive fragmentation and eventual disruption. As a result, the web-like network of filaments gradually dissipates over time, leaving behind a more diffuse and less organized ejecta distribution. At the age of $\sim 350$~years, the morphology of unshocked ejecta is that analyzed in Sect.~\ref{sec:res1}, recalling the morphology observed by JWST in \casa.

Our models also indicate that ejecta rich in intermediate-mass elements form ring- and crown-like structures in the shocked ejecta following interaction with the reverse shock (see \citealt{2021A&A...645A..66O}), resembling features observed in \casa\ (e.g., \citealt{2013ApJ...772..134M}). These structures, therefore, originate from the filamentary network of unshocked ejecta, more specifically from circular filaments surrounding the iron plumes. Formed as transient features, these rings and crowns gradually dissipate over time but persist in the shocked shell of our model at the age of \casa\ (see \citealt{2021A&A...645A..66O} for more detail). Their intricate morphology, such as the spiked appearance of the crowns, arises from fragmentation driven by HD instabilities. This suggests that these features may not remain visible for much longer, underscoring the importance of studying them in their present state to gain insights into the physical processes governing \casa's evolution.

\subsection{Future evolution of the network}
\label{sec:res3}

We extended our simulations to trace the evolution of the remnant up to an age of 1000~yr, aiming to investigate the future development of the filamentary structures in \casa\ and, more broadly, in core-collapse SNRs. In this way, we were able to estimate the longevity of these features and identify when they might stop to be observable. The lower panels in Fig.~\ref{evolution} illustrate the temporal evolution of the unshocked O-rich ejecta from now into the future, highlighting the persistence of filaments until their eventual disruption (see also Movie 2). According to the model, over the next centuries, the reverse shock will propagate deeper into the innermost ejecta, progressively disrupting the filamentary structures due to the growth of HD instabilities at the contact discontinuity. This process leads to the gradual destruction of the web-like network and erases the distinct imprint of the SN explosion.

In the specific simulation analyzed, the disruption process is accelerated by the presence of a dense circumstellar shell (\citealt{2022A&A...666A...2O}), which produces a second reflected shock. This shock is most energetic in regions where the shell density is highest, particularly in the northwest quadrant on the near side of the remnant. Consequently, the reverse shock propagates inward more rapidly in this region, increasing the asymmetry of its structure over time (see Fig.~4 in \citealt{2022A&A...666A...2O}). As the reverse shock advances, the ejecta in the western portion of the remnant are shocked earlier than those in the east. This asymmetry becomes evident by the time the remnant reaches an age of approximately $\approx 400$~yr (lower panels in Fig.~\ref{evolution}). By around 1000~yr, the reverse shock is highly asymmetric, having reached the explosion center from the western region, where it has heated all the ejecta and completely destroyed the network of filaments.

Our model predicts that the filaments will have been shock-dissipated and thus will become unobservable by roughly 700~yr after the explosion (see Movie 2). Beyond this point, the distribution of the presently unshocked ejecta transitions into a more mixed morphology, losing the intricate filamentary structures that characterize the younger remnant. This finding underscores the temporal fragility of these features and highlights the dynamic interplay of shocks and instabilities in shaping the observable properties of SNRs over time.

\section{Summary and conclusions}
\label{sec:conc}

We have investigated the origin and evolution of the intricate network of ejecta filaments recently unveiled in \casa\ by JWST, using high-resolution 3D HD and MHD simulations. Since the inclusion of magnetic fields did not significantly alter our results (see Appendix \ref{app:MHD}), we focused here on the HD simulations. By tracing the evolution of the remnant from the core-collapse SN explosion to an age of $\approx 1000$~years, our study sheds light on the physical processes that can shape a filamentary structure of unshocked ejecta in young SNRs.

Our simulations, based on a neutrino-driven SN explosion model, demonstrate that a web-like network of ejecta filaments forms naturally during the early phases of the remnant evolution, shaped by a combination of key processes. These include the expansion of neutrino-heated bubbles immediately following the core collapse, HD instabilities developing as the blast wave propagates through the stellar interior, and the Ni-bubble effect, which becomes significant shortly after the shock breakout. This latter effect enhances mixing and compresses lighter elements along the boundaries of initially Ni-rich domains, further contributing to the formation of a filamentary structure. The result is a dense, interconnected web of filaments enriched in O and other intermediate-mass elements (e.g., C, Ne, Mg, and Si).

We found that our models successfully reproduced the observed network of filaments of \casa\ with remarkable agreement in both morphology and complexity, offering new insights into the dynamic interplay of processes shaping SNRs and, possibly, providing a framework for interpreting analogous filamentary structures in other remnants. The key findings from our analysis can be summarized as follows.

\begin{itemize}
\item Filamentary structure at the age of \casa. The filamentary network, dominated by O-rich ejecta, forms an interconnected web-like structure that closely matches the morphology observed by JWST, although smaller length-scale features observed in \casa\ are not present, most likely due to the effect of numerical diffusion on features occupying a few cells. The modeled filaments exhibit distinct spatial organization, with intermediate-mass and light elements (e.g., C, O, Ne, and Mg) forming the network, while heavier elements (e.g., Ca, Ti, and Fe) are more centrally concentrated or confined to dense plumes. The filaments align systematically with channels on the surface of high-density Fe-rich ejecta, similar to rivers filling valleys. O-, Ne-, and Mg-rich ejecta permeate these channels, creating the intricate web-like pattern of ejecta observed by JWST. Large voids are present within the network, particularly in regions disrupted by extended Fe-rich plumes, consistent with ejecta layer inversions observed in the Fe-rich regions of \casa. The model accurately reproduces the observed filament thickness down to scales of $\sim 0.01$~pc, emphasizing the role of early explosion dynamics.\\

\item Formation of the filamentary network. According to our models, the network naturally originates from stochastic processes occurring shortly after core collapse, with neutrino-heated bubbles playing a central role in breaking the stratified layers of the progenitor star and driving turbulent mixing. Early expansion (during the first seconds following the core collapse) of Ni-rich bubbles or plumes compresses surrounding material rich in intermediate-mass elements such as O, creating dense sheaths of this material at the boundaries of these bubbles. HD instabilities, such as Rayleigh-Taylor instabilities, further shape the ejecta, enhancing the complexity of the network. The interaction between rising bubbles and adjacent ejecta layers results in the intricate topology observed in the filaments. In a few months following the shock breakout at the stellar surface (occurring $\approx 1500$~sec after core-collapse; \citealt{2017ApJ...842...13W}), the Ni-bubble effect due to the radioactive decay of $^{56}$Ni further enhances the filamentary structure, making the filaments thinner, brighter, and denser as a result of the expansion of matter in the Ni-rich domains, making the filaments more similar to those revealed with JWST observations.\\

\item Future evolution of the filamentary network. We extended the simulations beyond the age of \casa\ to make predictions on the future evolution of the network. We found that, over time, the interaction of the reverse shock with the ejecta fragments the filamentary structures due to the development of HD instabilities. By approximately 700~years after the core-collapse, the network becomes unobservable as the ejecta transitions to a more mixed morphology, erasing the distinct imprint of the explosion.
\end{itemize}

We note that the simulated network of ejecta filaments, while intricate, is not as complex or rich in fine structures as that observed by JWST in \casa\ (see \citealt{2024ApJ...965L..27M}, Dickinson et al. in preparation). Observational analysis suggests the presence of O-rich filaments extending into the innermost ejecta region, whereas in the model, the densest filaments predominantly concentrate in a shell enveloping the Fe-rich ejecta. A closer examination of the simulation revealed that O-rich ejecta are also present in the innermost region of the models, with velocities spanning a broad range from $-2500$ to $+4500$~km~s$^{-1}$. However, the density of these innermost O-rich features is significantly lower than that of the filaments concentrated in the shell. A detailed synthesis of emission in the bands of interest from the models, accounting for factors such as density, temperature, and other emission dependencies, could provide valuable insights into the detectability of innermost ejecta filaments.

Several factors could contribute to the discrepancies between models and observations. From a numerical point of view, numerical diffusion inherent even in high-resolution simulations can dampen small-scale features, smoothing gradients and limiting the development of finer structures (see Appendix \ref{app:resolution}). While the $2048^3$ resolution in run W15-IIb-sh-HD+dec-hr mitigates these effects to some extent, achieving even higher resolutions may be necessary to resolve the smallest-scale features and capture the full complexity of the ejecta especially in the innermost volume. Unfortunately, higher spatial resolution simulations would require a huge amount of numerical resources that makes these simulations very challenging.

Additionally, the spherically symmetric progenitor model used in the simulation may oversimplify the pre-collapse stellar interior. A more complex internal structure, including pre-existing turbulence, rotational effects, or instabilities (e.g., convective motions or shell-burning instabilities), could enhance the development of asymmetries during the explosion (e.g., \citealt{2011ApJ...733...78A, 2015ApJ...808L..21C, 2014ApJ...785...82S, 2015MNRAS.448.2141M}) and produce a more intricate filamentary network at shock breakout. A more sophisticated treatment of the physical processes occurring in the aftermath of core collapse and immediately following shock breakout could further enrich the network. These considerations suggest that both improvements in progenitor and SN-SNR modeling and higher-resolution simulations may produce an ejecta structure at the age of \casa\ closer to that observed.

Nevertheless, our analysis reveals that the intricate network of ejecta filaments discovered by JWST in \casa\ acts as a powerful archaeological record of the SN's earliest moments. By combining the results of these high angular resolution observations (\citealt{2024ApJ...965L..27M}, Dickinson et al. in preparation) with sophisticated MHD simulations, we have established a direct link between the observed filamentary structures and the fundamental processes governing SN evolution. More specifically, our models have shown that neutrino-driven explosions naturally generate complex filamentary ejecta networks through stochastic processes in the post-core-collapse phase, producing structures remarkably similar to those observed in \casa. The correspondence between model predictions and observations, therefore, indicates that these filaments serve as fossil records of the physical processes that dominated the initial phases of the explosion, providing a unique window into the mechanisms driving core-collapse SNe.

The filamentary network could also serve as a valuable diagnostic tool for investigating unresolved questions about \casa, particularly the origin of the S-rich "jets" that characterize the remnant's structure. Further observations of \casa's unshocked interior might provide crucial insights into their formation. For example, these jets might have left imprints or relic features in the distribution of unshocked Mg, Si, and Ca elements. Several scenarios could explain the origin of these jets: (a) they could be directly linked to the explosion mechanism; (b) they might result from post-explosion activity of a highly magnetized neutron star; (c) they could be relics of pre-collapse asymmetries (e.g., convective processes) in the progenitor star or (d) asymmetries in its CSM. 

Scenario (d) would become more plausible if the network of O-filaments in the unshocked interior shows no peculiarities in the jet/anti-jet directions. Conversely, peculiar features in these directions, such as particularly stretched or deformed filaments, could indicate scenario (c). If the jet and anti-jet directions appear to have been ``cleaned'' of O-rich filaments, showing radially extended voids, it would suggest scenarios (a) or (b), where the jets might have swept out matter, creating a more or less empty funnel in the filamentary network. A detailed 3D map of the filament network could be instrumental in addressing these questions, shedding light on whether the O-rich filament web exhibits directional asymmetries correlated with the jet and anti-jet. Such insights would be invaluable in disentangling the contributions of different physical processes to the jets' formation, clarifying their origin.

These findings pave the way for further studies of young SNRs, providing a robust framework for decoding the physical and chemical evolution of stellar explosions. This study also underscores the transformative potential of high angular resolution observations, such as those from JWST, in uncovering fundamental physical processes during SN explosions. Equally crucial is the integration of such observations with advanced numerical simulations, offering a powerful approach to interpreting the intricate legacy of SN explosions in young remnants like \casa. Expanding this analysis to a broader sample of young SNRs using JWST’s capabilities holds immense promise for advancing our understanding of the multi-scale physics governing the final fate of massive stars.

\begin{acknowledgements}
We thank an anonymous referee for the useful suggestions that allowed us to improve the manuscript. S.O. expresses sincere gratitude to Massimiliano Guarrasi (CINECA) for his invaluable support in using the high performance computing (HPC) facilities at CINECA: the high-resolution simulations would not have been possible without his assistance. The \PLUTO\ code is developed at the Turin Astronomical Observatory (Italy) in collaboration with the Department of General Physics of  Turin University (Italy) and the SCAI Department of CINECA (Italy). We acknowledge the CINECA ISCRA Award N.HP10BUMIQR for the availability of HPC resources and support at the infrastructure Leonardo based in Italy at CINECA. Additional computations were carried out on the HPC system MEUSA at the SCAN (Sistema di Calcolo per l'Astrofisica Numerica) facility for HPC at INAF-Osservatorio Astronomico di Palermo. Computer resources for this project have also been provided by the Max Planck Computing and Data Facility (MPCDF) on the HPC systems Cobra and Draco. The navigable 3D graphics have been developed in the framework of the project 3DMAP-VR (3-Dimensional Modeling of Astrophysical Phenomena in Virtual Reality; \citealt{2019RNAAS...3..176O, 2023MmSAI..94a..13O}) at INAF-Osservatorio Astronomico di Palermo.
S.O., M.M., and F.B. acknowledge financial contribution from the PRIN 2022 (20224MNC5A) - ``Life, death and after-death of massive stars'' funded by European Union – Next Generation EU, and the INAF Theory Grant ``Supernova remnants as probes for the structure and mass-loss history of the progenitor systems''. H.-T.J.\ acknowledges support by the German Research Foundation (DFG) through the Collaborative Research Centre ``Neutrinos and Dark Matter in Astro- and Particle Physics (NDM),'' grant No. SFB-1258-283604770, and under Germany's Excellence Strategy through the Cluster of Excellence ORIGINS EXC-2094-390783311. I.D.L. acknowledges funding from the Belgian Science Policy Office (BELSPO) through the PRODEX project ``JWST/MIRI Science exploitation'' (C4000142239) and funding from the European Research Council (ERC) under the European Union's Horizon 2020 research and innovation program DustOrigin (ERC-2019-StG-851622).
T.T. acknowledges support from the NSF grant AST-2205314 and the NASA ADAP award 80NSSC23K1130.
D.J.P. acknowledges support from the Chandra X-ray Center, which is operated by the Smithsonian Institution under NASA contract NAS8-03060.
\end{acknowledgements}

\bibliographystyle{aa}
\bibliography{references}

\clearpage
\begin{appendix}
\onecolumn

\section{Effect of spatial resolution}
\label{app:resolution}

\begin{figure*}[!hb]
   \centering
   \includegraphics[width=\textwidth]{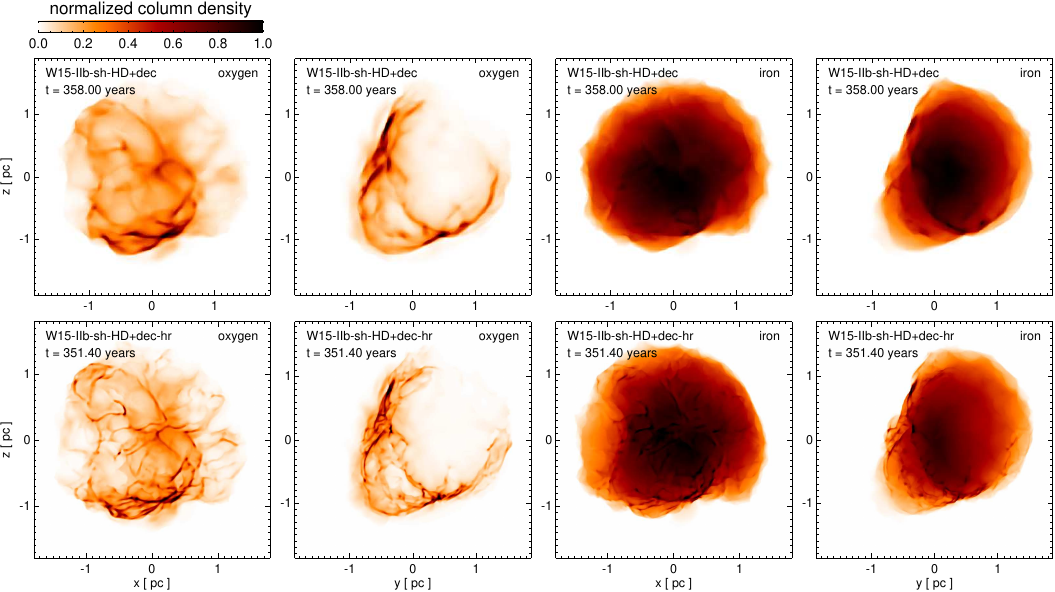}
   \caption{Comparison of two simulations differing only in spatial resolution (see Table~\ref{Tab:model}): run W15-IIb-sh-HD+dec, calculated on a grid with $1024^3$ grid points (upper panels), and run W15-IIb-sh-HD+dec-hr, calculated on a grid with $2048^3$ grid points (lower panels). The figure presents volumetric renderings of the mass density distribution for unshocked O-rich (first two columns on the left) and Fe-rich (last two columns on the right) ejecta, integrated along the LoS, at the age of \casa. Each image is normalized to its maximum for better visibility (color bar on the top left of the figure). The first and third columns display the volumetric renderings as viewed from Earth's perspective, corresponding to the negative $y$-axis (front view). The second and last columns show the renderings from a vantage point along the positive $x$-axis (side view).}
   \label{fig_resolution}%
   \end{figure*}

We investigated the impact of spatial resolution on the ability to accurately capture the intricate structures of SN ejecta by comparing two simulations differing only in their resolution (see Table~\ref{Tab:model}): run W15-IIb-sh-HD+dec, performed on a grid with $1024^3$ grid points (upper panels), and run W15-IIb-sh-HD+dec-hr, performed on a grid with $2048^3$ grid points (lower panels). In both simulations, a network of ejecta filaments is evident (see Fig.~\ref{fig_resolution}), confirming that our simulations have enough spatial resolution to capture the large-scale structure of unshocked ejecta. However, in the lower-resolution run (W15-IIb-sh-HD+dec), the unshocked O- and Fe-rich ejecta appear smoother, with fewer discernible small-scale features. By contrast, the higher-resolution simulation (W15-IIb-sh-HD+dec-hr) unveils a much more detailed and complex morphology. The filamentary structures of the ejecta, especially in the O-rich material, are sharper, more interconnected, and exhibit finer-scale variations not visible in the lower-resolution counterpart. This improvement highlights the higher-resolution simulation's capacity to more accurately resolve the post-SN dynamics and evolution of ejecta following the shock breakout at the stellar surface.

The differences observed between the two simulations considered can largely be attributed to the effects of numerical diffusion, which tends to smooth sharp gradients and suppress small-scale structures in lower-resolution models. In the $1024^3$ simulation, numerical diffusion likely smooths the formation of thin, dense filaments. Consequently, the ejecta morphology is less representative of the physical processes shaping the ejecta in early phases of evolution. In the $2048^3$ simulation, the reduced impact of numerical diffusion allows a better description of the Ni-bubble effect, which compresses filamentary structures in the months following the shock breakout (see Sect.~\ref{sec:res2}), making them thinner and denser. This results in a filamentary network that more closely resembles the intricate structures observed in \casa. However, since the filaments in the high-resolution run are resolved by only $\approx 10$ grid points, it is possible that some degree of numerical diffusion persists, and even finer structures might emerge with further increases in spatial resolution. 

For a given simulation, spatial resolution can impact the structures that develop at different distances $r$ from the explosion center. The effective spatial resolution at a given $r$ can be estimated as $\delta r/r$, where $\delta r$ represents the local cell size. In Cartesian coordinates, $\delta r$ varies depending on the angle of the unit vector $\hat{r}$ relative to the grid axes, with the minimum possible value taken as the reference resolution. At the age of \casa, most of the thin and bright filaments in run W15-IIb-sh-HD+dec-hr are located between 0.5 and 1 pc, where the effective spatial resolution $\delta r/r$ ranges from approximately 0.0025 to 0.005. In contrast, structures in the innermost regions are resolved with lower spatial resolution ($\delta r/r>0.005$). This resolution limitation was even more pronounced in the early phase of the remnant’s evolution when the filamentary network was still confined to the innermost regions. However, as shown above, we assessed the impact of resolution by doubling the number of grid points, confirming that our high-resolution simulations reliably capture the main features of the filamentary structures. As the ejecta expand to larger radii, their resolution improves (as $\delta r/r$ decreases), benefiting from improved grid resolution in the outer regions. This ensures that the essential physical processes governing the evolution of the ejecta are accurately resolved throughout the simulation.

To better capture these small-scale structures, future simulations would require either an adaptive mesh refinement (AMR) approach or a globally higher resolution, particularly in the remnant's interior, where the filaments initially form and evolve. These findings underscore the importance of using high-resolution simulations to study the fine-scale features of SNRs and to achieve reliable comparisons with high-precision observational data, such as those obtained with JWST.

\section{Effects of the magnetic field on the structure of unshocked ejecta}
\label{app:MHD}

We investigated the effects of the magnetic field on the structure and dynamics of unshocked ejecta, by analyzing run W15-IIb-sh-MHD+dec-hr using the same approach as for run W15-IIb-sh-HD+dec-hr. Our findings confirm that the presence of a magnetic field does not significantly affect the structure or dynamics of the unshocked ejecta, reinforcing the results obtained for run W15-IIb-sh-HD+dec-hr.

\begin{figure*}[!hb]
   \centering
   \includegraphics[width=\textwidth]{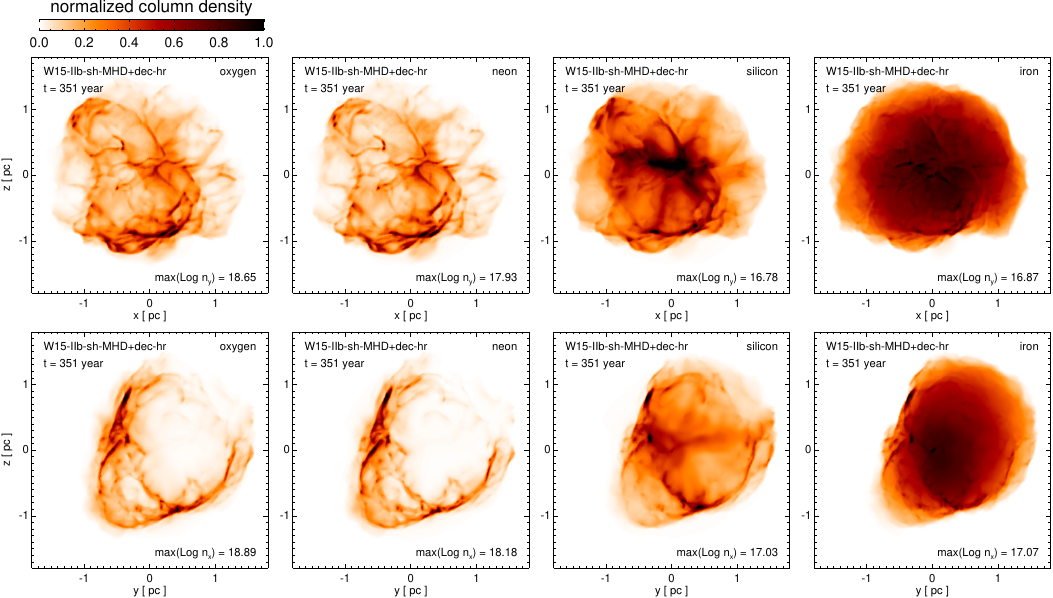}
   \caption{Same as Figs.~\ref{maps_all_elem_front} and \ref{maps_all_elem_side}, but for run W15-IIb-sh-MHD+dec-hr, showing the distributions of O, Ne, Si, and Fe. The vantage point is along the negative $y$-axis (upper panels) and the positive $x$-axis (lower panels).}
   \label{maps_MHD}%
   \end{figure*}

As an example, Fig.~\ref{maps_MHD} presents 3D volumetric renderings of the unshocked ejecta distribution across different species, observed from two distinct viewing angles, obtained from run W15-IIb-sh-MHD+dec-hr. These figures are analogous to Figs.~\ref{maps_all_elem_front} and \ref{maps_all_elem_side} for run W15-IIb-sh-HD+dec-hr. The differences between the MHD and HD cases are minimal. In particular, the ejecta distributions in the MHD case appear slightly smoother due to the increased numerical diffusivity of the MHD solver at the same spatial resolution. Since this study focuses on the smallest, sharpest features of the ejecta distribution (best resolved in the HD simulation) we presented the results of run W15-IIb-sh-HD+dec-hr in the main body of the paper.

While the magnetic field has little influence on unshocked ejecta, it plays a crucial role in the dynamics of shocked ejecta. In particular, it suppresses the growth of HD instabilities (such as Rayleigh-Taylor and Kelvin-Helmholtz instabilities) at the contact discontinuity between shocked ejecta and the shocked CSM and can enhance the efficiency of radiative cooling (e.g., \citealt{2008ApJ...678..274O}). In a companion paper (\citealt{Orlando2025}), we investigate the structure of a shocked, dense CSM shell in front of \casa\ recently explored with JWST and termed the  ``Green Monster'' (\citealt{2024ApJ...965L..27M, 2024ApJ...976L...4D}). In this separate work we examine in detail the role of the magnetic field in shaping newly discovered hole and ring features and, more generally, the structure of the mixing region between the forward and reverse shocks.

\section{Online multi-media material}
\label{app:multi-media}

The present paper is accompanied by a collection of online movies and interactive 3D graphics that visually capture the intricate structure and dynamic evolution of the web-like network of ejecta filaments. These supplementary materials are designed to enhance understanding by exploring the ejecta morphology and its temporal progression, offering a detailed perspective on the phenomena discussed in the study. Below, we provide a comprehensive description of these resources.

\medskip
\noindent
{\bf Online Movies.} These movies are available as supplementary material on the A\&A webpage:

\begin{itemize}

\item Movie 1: "Virtual Tour Through Unshocked Ejecta". This movie complements Fig.~\ref{ejecta_structure}, providing a guided virtual tour through the unshocked ejecta regions rich in Fe and O. It highlights the complex spatial arrangement of ejecta filaments within the remnant interior.

\item Movie 2: "Evolution of Ejecta Composition Over 900 Years". This movie complements Fig.~\ref{sn_evolution} and presents the full evolution of the unshocked ejecta. The left panel focuses on O-rich material, while the right panel tracks the Ni/Fe-rich ejecta, from the initial explosion at $t = 1.3$~s to the remnant stage at $t = 900$~years. This animation captures the effects of physical processes at work during the complex phases of SN evolution and the interplay between ejecta components.

\end{itemize}

\noindent
{\bf Interactive 3D Graphics.} These models are available as supplementary material on the A\&A webpage and accessible online through the sketchfab website, offering an interactive experience - even through virtual reality device (3DMAP-VR project; \citealt{2019RNAAS...3..176O, 2023MmSAI..94a..13O}) - to complement the figures in the paper:

\begin{itemize}
\item Model 1: "A Network of Ejecta Filaments in Cassiopeia A". Accessible at \url{https://skfb.ly/psXKr}, this model complements Fig.~\ref{ejecta_structure}, enabling users to interactively explore the distribution of Fe-rich and O-rich ejecta. Features include zooming, panning, and rotating views, as well as labeled annotations that highlight key elements such as: (1) Dense, Fe-rich ejecta; (2) Filaments of O-rich ejecta; (3) The position of the reverse shock.

\item Model 2: "Ejecta Structure at Shock Breakout". Accessible at \url{https://skfb.ly/psXKs}, this model complements Fig.~\ref{sh_breakout} and focuses on the initial conditions and the distribution of ejecta shortly after the shock breakout. Labeled annotations provide details on: (1) Plumes of Ni-rich ejecta; and (2) Filaments of O-rich ejecta.
\end{itemize}

These materials are integral to a deeper understanding of the study, offering interactive and dynamic representations of the phenomena discussed in the paper. We encourage readers to explore these resources for a more comprehensive grasp of the results.

\end{appendix}

\end{document}